\documentclass[pre,aps,twocolumn,showpacs,floatfix,10pt]{revtex4-1}
\usepackage{graphicx}
\usepackage{color}
\usepackage{amsmath, amsthm, amssymb}
\newcommand{\be}{\begin{equation}}
\newcommand{\ee}{\end{equation}}
\newcommand{\bea}{\begin{eqnarray}}
\newcommand{\eea}{\end{eqnarray}}

\begin{document}

\title{Asymptotic Behavior of the Isotropic-Nematic and Nematic-Columnar Phase Boundaries for the System of Hard Rectangles on a Square lattice}
\author{Joyjit Kundu}
\email{joyjit@imsc.res.in}
\affiliation{The Institute of Mathematical Sciences, C.I.T. Campus,
Taramani, Chennai 600113, India}
\author{R. Rajesh}
\email{rrajesh@imsc.res.in}
\affiliation{The Institute of Mathematical Sciences, C.I.T. Campus,
Taramani, Chennai 600113, India}

\date{\today}

\begin{abstract}

A system of hard rectangles of size $m\times mk$ 
on a square lattice undergoes three entropy driven phase 
transitions with increasing density for large enough aspect ratio $k$: first 
from a low density isotropic to an intermediate density nematic phase, 
second from the nematic to a columnar phase, and third from the columnar 
to a high density sublattice phase. 
In this paper we show,
from extensive Monte Carlo simulations of systems with
$m=1,2$ and $3$,  that the transition density for 
the isotropic-nematic transition is $\approx A_1/k$ when $k \gg 1$, 
where $A_1$ is independent of $m$. We estimate $A_1=4.80\pm 0.05$.
Within a Bethe approximation, we obtain $A_1=2$ and the virial expansion
truncated at second virial coefficient gives
$A_1=1$. The critical density for the 
nematic-columnar transition when $m=2$ is numerically shown to
tend to a value less than the full packing density as $k^{-1}$ when 
$k\to \infty$. We find that the critical Binder cumulant for  this 
transition is non-universal
and decreases as $k^{-1}$ for $k \gg 1$.
However, the transition is shown to be in the
Ising universality class.
\end{abstract}


\maketitle

\section{\label{sec:intro}Introduction}

Hard core lattice gas models of particles interacting only through 
excluded volume interaction continue to be of interest in Statistical 
Physics. They are minimal models to study entropy driven phase 
transitions, have direct realizations in adsorption of gas particles on 
metal 
surfaces~\cite{taylor1985,bak1985,rikvold1991,koper1998,binder2000,evans2000}, 
and are closely related to the freezing 
transition~\cite{alder1957,alder1962}, directed and undirected 
animals~\cite{deepak2,deepak1,imbrie} and the Yang-Lee 
singularity~\cite{parisi}.  Systems of differently shaped particles on 
different lattices have been studied both analytically and numerically. 
Examples include 
squares~\cite{gaunt1965,nigam1967,pearce1988,baram1994,nienhuis2011,kabir2012}, 
hexagons~\cite{baxter1980,heilmann1973}, dimers with nearest neighbor 
exclusion~\cite{dickman2012}, triangles~\cite{nienhuis1999}, 
tetrominoes~\cite{barnes2009}, rods~\cite{ghosh2007,joyjit2013}, 
rectangles~\cite{joyjit_rectangle}, lattice models for 
discs~\cite{fernandes2007,trisha_knn} and mixtures~\cite{kabir2014}.

In this paper, we focus on the system of hard rectangles of size $m 
\times m k$ on the square lattice, where each rectangle occupies $m$ 
sites along the short axis and $m k$ sites along the long axis, where 
$k$ is the aspect ratio, and $m$, $k$ are integers. In recent times, 
there has been renewed interest in this problem when, even though both 
the low density and the maximal density phases are known to be 
disordered~\cite{degennesBook}, the existence of a nematic phase for the 
hard rod ($m=1$) system was convincingly demonstrated in simulations for 
$k \geq 7$~\cite{ghosh2007}, and thereafter proven rigorously for $k \gg 
1$~\cite{giuliani2013}.  The maximal density phase being disordered 
implies the existence of a second entropy driven transition with 
increasing density. This has been established in 
simulations~\cite{joyjit2013}. While the first transition into the 
nematic phase is in the Ising universality class~\cite{fernandez2008a}, 
the universality class of the second transition into the high density 
disordered phase is not that clear~\cite{joyjit2013,joyjit_rltl2013}.

The phase diagram for rectangles with $m>1$ is even richer, and was 
recently determined using Monte Carlo simulations for $m=2,3$ and $k\leq 
7$, and generalized to larger $m,k$ using entropic 
arguments~\cite{joyjit_rectangle}. For $k\geq 7$, with increasing 
density, the system transits successively from isotropic (I) to nematic 
(N) to columnar (C) to solid-like sublattice (S) phases (see 
Sec.~\ref{sec:model} for a precise define of the phases). When $k<7$, 
the $N$ phase is absent and the system makes a direct transition from 
the $I$ phase to a $C$ phase, if present. The C phase is absent only for 
$m=2$ and $k=2,3$, and the system makes a direct transition from the I 
phase into the $S$ phase. A detailed study of the nature of the 
transitions may be found in Ref.~\cite{joyjit_rectangle}.

In this paper, we focus on the asymptotic behavior of the 
isotropic-nematic (I-N) and nematic-columnar (N-C) phase boundaries for 
large aspect ratio $k$. It was heuristically argued in 
Ref.~\cite{joyjit_rectangle} that the limit $k\to \infty$ keeping $m$ 
fixed should correspond to the limit of oriented lines in the continuum, 
and thus the critical density should be independent of $m$. For this 
limiting case in three dimensions, the virial expansion truncated at the 
second virial coefficient becomes exact and the critical density for the 
I-N transition $\rho_c^{I-N} \approx A_1 
k^{-1}$~\cite{onsager1949,zwanzig1963,vroege1992}. $A_1$ for oriented 
long rectangles in the two-dimensional continuum can be directly estimated by 
simulating oriented lines of length $\ell$, for which it is 
straightforward to show that the critical number density $\approx A_1 
\ell^{-2}$. From the simulations of this system with $\ell=1$, it can be 
inferred that $A_1 \approx 4.84$~\cite{fischer2009}. For $m=1$, by 
simulating systems with $k$ up to $12$ on the lattice, it has been shown 
that $\rho_c^{I-N} \propto k^{-1}$~\cite{fernandez2008c}. From the value 
of the critical density for $k=10$~\cite{fernandez2008c}, it can be 
estimated that $A_1\approx 5.02$, different from that for oriented 
lines~\cite{fischer2009}.  There are no such similar studies for $m>1$.

For the N-C transition, the limit $k\to \infty$ keeping $m$ fixed, 
corresponds to asking whether oriented long rectangles in the continuum 
exhibit a C phase. The N-C transition has been studied using a Bethe 
approximation that predicts the critical density to be $\rho_c^{N-C} 
\approx A_2(m)+A_3(m) k^{-1}$ when $k\to \infty$, where $A_2(m) \leq 
1$~\cite{joyjit_rectangle}. The approximations being ad-hoc, it is 
important to have numerical confirmation of these results, but none 
exists. The other limit $m \to \infty$, keeping $k$ fixed corresponds to 
the continuum problem of oriented rectangles of aspect ratio $k$, a 
model that was introduced and studied by Zwanzig using virial 
expansion~\cite{zwanzig1963}.  This limit is difficult to study 
numerically on the square lattice.

In this paper, by simulating systems of rectangles with aspect ratio $k$ 
up to $60$ (for $m=1$), and $k=56$ (for $m=2$ and $3$), we show that 
$\rho_c^{I-N}$ is proportional to $k^{-1}$ for $m=1,2$ and $3$. Within 
numerical error, $A_1$ is shown to be independent of $m$ and equal to 
$4.80 \pm 0.05$. To understand better the limit of large $k$, we study 
the I-N transition using a Bethe approximation, and a virial expansion 
truncated at the second virial coefficient. The critical density 
$\rho_c^{I-N}$ is obtained for all $m$ and $k$. For large $k$, both 
theories predict that $\rho_c^{I-N}$ is independent of $m$, while 
$A_1=2$ in the Bethe approximation and $A_1=1$ in the truncated virial 
theory. For the N-C transition, we numerically determine 
$\rho_c^{N-C}$ for $m=2$ and $k$ up to $24$. We show that for large $k$, 
$\rho_c^{N-C} \approx 0.73+0.23 k^{-1}$, consistent with the 
calculations in Ref.~\cite{joyjit_rectangle}. This shows that a system 
of oriented rectangles with large aspect ratio in the two dimensional 
continuum should exhibit both nematic and columnar phases. In addition, 
we find that the Binder cumulant at the N-C transition is surprisingly 
dependent on $k$, and decreases as $k^{-1}$ with increasing $k$. However, 
we show 
that the transition remains in the Ising universality class.

The rest of the paper is organized as follows. Section~\ref{sec:model} 
contains a definition of the model, a brief description of the Monte 
Carlo algorithm, and a definition of the phases and the relevant 
thermodynamic quantities of interest. In Sec.~\ref{sec:phase_bound_IN}, 
we present the numerical results for the I-N transition for $m=1,2$ and $3$. 
In Sec.~\ref{sec:phase_bound_NC}, we numerically determine the 
asymptotic behavior of the N-C phase boundary for $m=2$ and large $k$. 
The Binder cumulant is shown to be non-universal, though exponents 
continue to be universal. Section~\ref{sec:IN_MF} contains calculations 
of the I-N phase boundary using an ad-hoc Bethe approximation and a 
truncated virial expansion. Section~\ref{sec:summary} contains a summary 
and discussion of results.

\section{\label{sec:model}Model and Definitions}

We consider a system of monodispersed hard rectangles of size $m \times 
m k$ on a square lattice of size $L \times L$, with periodic boundary 
conditions. Each rectangle occupies $m$ sites along the short axis and 
$m k$ sites along the long axis, such that $k$ is the aspect ratio. A 
rectangle is called horizontal or vertical depending on whether the long 
axis is along the x-axis or y-axis. No two rectangles may overlap, or 
equivalently a lattice site may be occupied by utmost one rectangle. We 
associate an activity $z=e^{\mu}$ to each rectangle, where $\mu$ is the 
chemical potential.

We simulate the system in the constant $\mu$ grand canonical ensemble 
using an efficient algorithm that involves cluster moves.  The 
implementation of the algorithm for the system of hard rectangles is 
described in detail in Ref.~\cite{joyjit_rectangle}. The algorithm has 
been shown to be very useful in equilibrating hard core systems of 
extended particles at high densities. Other implementations of the 
algorithm include lattice models of hard 
rods~\cite{joyjit_dae,joyjit2013} and hard discs~\cite{trisha_knn} and 
mixtures of dimers and hard squares~\cite{kabir2014}.

The data presented in the paper corresponds to systems with aspect ratio 
up to $60$ and system sizes up to $L = 1680$. In the simulations, we 
ensured equilibration by confirming that the final state is independent 
of the initial configuration. The system is typically equilibrated for 
$10^7$ Monte Carlo steps, and the measurement is broken into $10$ 
statistically independent blocks, each of size $10^7$ Monte Carlo steps.

The system of hard rectangles may exist in one of the four different 
phases: isotropic (I), nematic (N), columnar (C) and sublattice 
(S)~\cite{joyjit_rectangle}. In the low density I phase the system 
neither possess orientational order nor positional order. The N phase 
breaks the orientational symmetry by preferring a particular 
orientation, either horizontal or vertical. However, the N phase has no 
positional order. In the C phase, along with orientational order, the 
system possesses partial positional order in the direction perpendicular 
to the preferred orientation. In the C phase of $2\times 2k$ rectangles, 
if most of the rectangles are horizontal (vertical), then the heads 
(bottom left corner) of the rectangles mostly occupy either even or odd 
rows (columns). The high density S phase has positional order along both 
horizontal and vertical directions, but no orientational order. In this 
phase the heads of the rectangles preferentially occupy one of $m^2$ 
sublattices~\cite{joyjit_rectangle}.

We now define the order parameters and relevant thermodynamic 
quantities used to study the I-N and 
the N-C transitions. For the I-N transition, we define a order parameter 
\be
Q_1 = \frac{\langle |N_h-N_v| \rangle}{\langle N_h+N_v \rangle}, 
\label{eq:q1}
\ee
where $N_h$ and $N_v$ are the number of horizontal and vertical 
rectangles respectively. $Q_1$ is zero in the I phase and nonzero in the 
N and C phases. 
For the N-C transition, we define an order parameter only for $m=2$ as our 
simulations are restricted to this value of $m$. Generalization
to larger $m$ is straightforward~\cite{joyjit_rectangle}. Let
\be
Q_2 = \frac{\langle ||N_{re}-N_{ro}| -|N_{ce}-N_{co}||\rangle}{\langle 
N_h+N_v \rangle},
\ee
where $N_{re}$ ($N_{ro}$) is the number of rectangles whose heads are in the 
even (odd) rows, 
and $N_{ce}$ ($N_{co}$) is the number of rectangles whose heads are in the 
even (odd) columns.  In the I and N phases, $N_{re} \approx N_{ro}$, and 
$N_{ce} \approx N_{co}$, and hence $Q_2$ is zero. In the C phase, either
$N_{re} \neq N_{ro}$ and  $N_{ce} \approx N_{co}$, or 
$N_{re}  \approx N_{ro}$ and  $N_{ce} \neq N_{co}$, such that $Q_2$ is 
non-zero.

The second moment 
of the order parameter $\chi_i$ and the Binder cumulant $U_i$ are defined as,
\begin{subequations}
\label{eq:thermo-definition}
 \bea
\chi_i&=&\langle Q_i^2 \rangle L^2, \label{eq:chi}\\
U_i&=&1- \frac{\langle Q_i^4 \rangle} {3 \langle Q_i^2 \rangle ^2},
\label{eq:U}
\eea
\end{subequations}
where $i=1,2$.
The thermodynamic quantities become singular at the transition. Let 
$\epsilon=(\mu-\mu_c)/\mu_c$, where $\mu_c$ is the critical chemical 
potential. The singular behavior is characterized by the critical 
exponents $\beta$, $\gamma$, $\nu$ defined by $Q_i \sim 
(-\epsilon)^\beta$, $\epsilon<0$, $\chi_i \sim
|\epsilon|^{-\gamma}$, and $\xi_i
\sim |\epsilon|^{-\nu}$, where $\xi_i$ is the correlation length, 
$|\epsilon| \rightarrow 0$, and $i=1,2$. 
The critical exponents may be obtained numerically through
finite size scaling near the critical point:
\begin{subequations}
\label{eq:scaling}
\bea
U_i &\simeq& f_u(\epsilon L^{1/\nu}), \label{eq:Uscaling}\\
Q_i &\simeq& L^{-\beta/\nu} f_q(\epsilon L^{1/\nu}), \label{eq:Qscaling}\\
\chi_i & \simeq& L^{\gamma/\nu} f_{\chi}(\epsilon L^{1/\nu}),
\label{eq:chiscaling}
\eea
\end{subequations}
where $f_u$, $f_q$, and $f_{\chi}$ are scaling functions. 

\section{\label{sec:phase_bound_IN}Asymptotic behavior of the 
Isotropic--Nematic phase boundary: Numerical Study}

\begin{figure}
\includegraphics[width=\columnwidth]{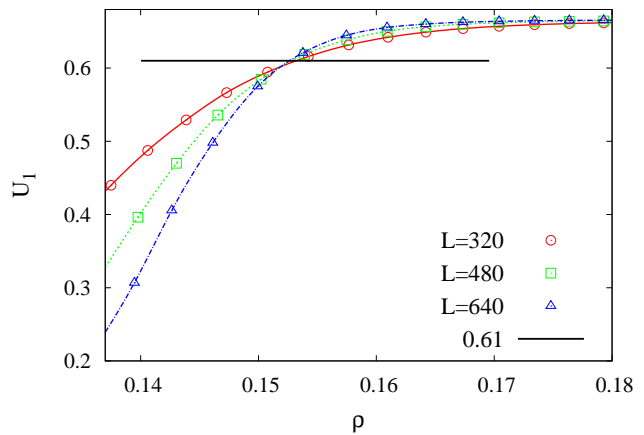}
\caption{(Color online) The variation of the Binder cumulant $U_1$ with density 
$\rho$ for three different system sizes. The lines are cubic splines,
fitted to the data.
The value of $U$ at $\rho=\rho_c$ is $\approx 0.61$. The data are for $m=1$ and $k=32$.}
\label{fig:bc_m1_IN} 
\end{figure}

In this section we investigate the asymptotic behavior of the I-N phase 
boundary for $m=1$, $2$, and $3$ by numerical simulations and show
that $\rho_c^{I-N} = A_1 k^{-1}$ when $k\gg 1$, where $A_1$ is independent 
of $m$. Since 
there are two symmetric N phases (horizontal and vertical), the I-N 
transition for the system of hard rectangles is continuous and belongs 
to the Ising universality class for all $m$~\cite{joyjit_rectangle}. We 
determine the critical density $\rho_c^{I-N}$ from the point of 
intersection of the curves of Binder cumulant with density for different 
system sizes. A typical example is shown in Fig.~\ref{fig:bc_m1_IN}, 
where the variation of $U_1$ with density $\rho$ is shown for three 
different system sizes when $m=1$ and $k=32$. The Binder cumulant data 
are fitted to a cubic spline to obtain a smooth and continuous curve for 
each $L$. This allows us to determine the point of intersection or 
$\rho_c^{I-N}$ more accurately. In the example shown in 
Fig.~\ref{fig:bc_m1_IN}, the curves for Binder cumulants for three 
different system sizes crosses at $\rho=\rho_c^{I-N}\approx 0.152$ and 
the value of the critical Binder cumulant $U_1^c \approx 0.61$. We find 
$U_1^c \approx 0.61$ for all values of $m$ and $k$ that we have studied, 
consistent with the value for the two dimensional Ising 
model~\cite{blote1993}.

\begin{figure}
\includegraphics[width=\columnwidth]{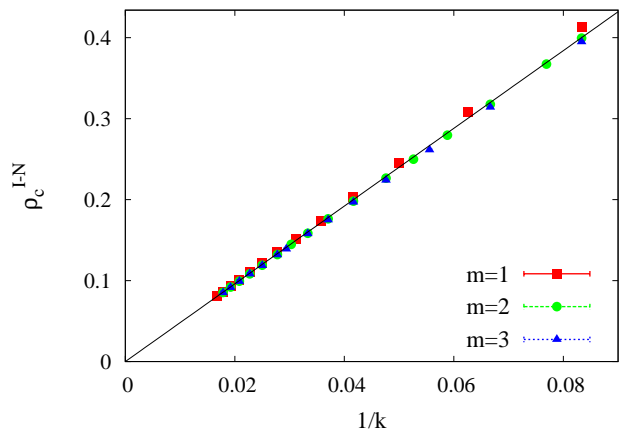}
\caption{(Color online) The variation of the critical density for the I-N transition 
$\rho_c^{I-N}$ with $k^{-1}$ for $m=1,2$ and $3$.
The straight line is $4.80 k^{-1}$.}
\label{fig:rhoc_IN} 
\end{figure}

We  simulate systems with 
aspect ratio up to $k=60$ for $m=1$ and $k=56$ for $m=2$ and $3$. 
The critical density $\rho_c^{I-N}$ obtained from the Binder cumulants
are shown in Fig.~\ref{fig:rhoc_IN}. The data are clearly linear in $k^{-1}$ for large $k$, 
confirming that $\rho_c^{I-N} = A_1 k^{-1}$, $k\gg 1$.
In addition, the data for $m=1,2,3$ asymptotically lie on the same straight line,
showing that $A_1$ is independent of $m$. We estimate 
$A_1 = 4.80 \pm 0.05$. 

\section{\label{sec:phase_bound_NC}Asymptotic behavior of the Nematic--Columnar 
phase boundary: Numerical Study}

In this section, we numerically study the N-C phase transition for $m=2$ and 
determine the asymptotic behavior of the critical density 
$\rho_c^{N-C}$ for large $k$. 
When $m=2$, the N-C transition belongs to the Ising universality class for all $k$ and 
the corresponding critical densities are determined from the intersection of 
Binder cumulant curves for  different system sizes as discussed in 
Sec.~\ref{sec:phase_bound_IN}. 

\begin{figure}
\includegraphics[width=\columnwidth]{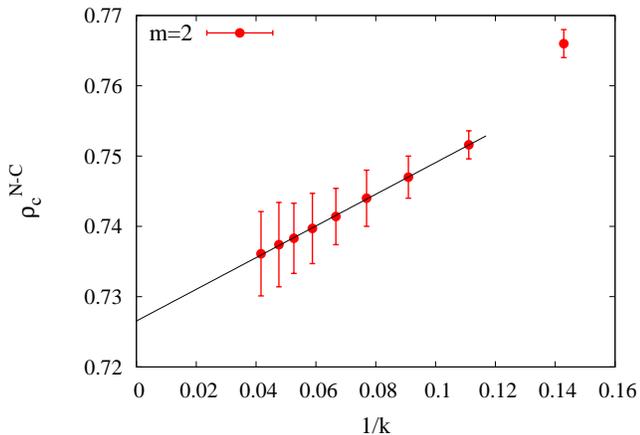}
\caption{(Color online) The variation of the critical density for the 
N-C transition $\rho_c^{N-C}$ with $k^{-1}$ for $m=2$. The straight line 
is a linear fit to the data: $0.727+0.226 k^{-1}$.}
\label{fig:rhoc_NC} 
\end{figure}

The critical density $\rho_c^{N-C}$ decreases to a constant with increasing $k$ 
(see Fig.~\ref{fig:rhoc_NC}). We obtain $\rho_c^{N-C} \approx 0.73+0.23 k^{-1}$,
$k \gg 1$ when $m=2$. These results are in qualitative agreement with the 
predictions of the Bethe
approximation: $\rho_c^{N-C} \approx A_2(m)+B_2(m) k^{-1}$, for $k \gg 1$. 
Within the Bethe approximation $A_2 \approx 0.59$ and 
$B_2 \approx 0.15$ for $m=2$~\cite{joyjit_rectangle}. 
As $\rho_c^{N-C}$ asymptotically approaches a constant value, it becomes increasingly 
difficult to get reliable data for large $k$. 

\begin{figure}
\includegraphics[width=\columnwidth]{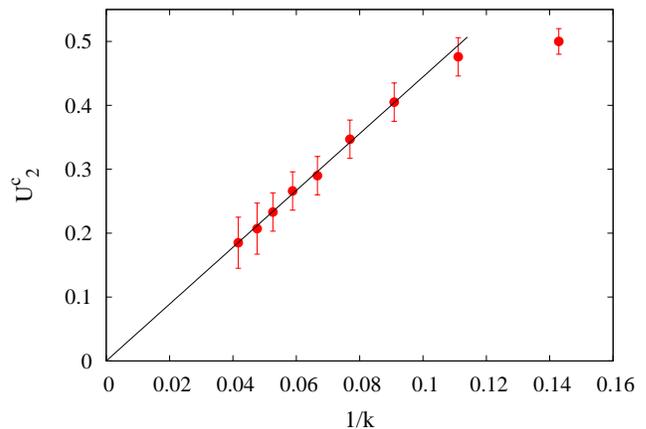}
\caption{(Color online) The variation of the critical Binder cumulant 
$U_2^c$ at the N-C transition with $k^{-1}$. 
The straight line $4.446 k^{-1}$ is a linear fit to the data. 
The data are for $m=2$.}
\label{fig:bc_k} 
\end{figure}

Surprisingly, we find that the value of the critical Binder cumulant at the N-C 
transition point depends on the 
aspect ratio $k$. When $m=2$, the critical Binder cumulant $U^c_2$ 
decreases monotonically as a power law 
with $k$, from $0.50$ when $k=7$ to 
$0.18$ when $k=24$ (see Fig.~\ref{fig:bc_k}). The data
is fitted best with $U^c_2 \approx 4.45 k^{-1}$.
Usually, for the Ising universality class, the value critical Binder cumulant 
at the transition point is expected to be universal ($\approx 0.61$) . 
However, there are a few examples of systems that exhibit such non universal 
behavior~\cite{blote1993,selke2005,selke2007}.  These include
the anisotropic Ising model where the critical Binder cumulant depends on
the ratio of the coupling constants along the $x$ and $y$ 
directions~\cite{selke2005},
and the isotropic Ising model  on rectangular lattice, where the critical
Binder cumulant is a function of the aspect ratio of the underlying 
lattice~\cite{blote1993}. In the latter case, $U^c_2
\approx 2.46 \alpha^{-1}$, where $\alpha$ is the aspect
ratio of the lattice~\cite{blote1993}. Thus, nominally $k \approx 1.8 \alpha$. 

\begin{figure}
\includegraphics[width=\columnwidth]{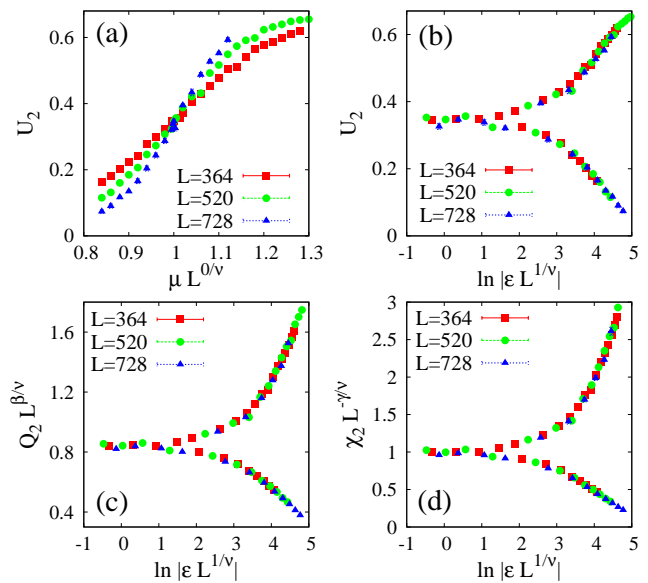}
\caption{(Color online) The data for different $L$ 
near the N-C transition collapse when scaled with the
Ising exponents $\beta/\nu=1/8$, $\gamma/\nu=7/4$, $\nu = 1 $ and $\mu_c \approx 1.00$. 
Data are for rectangles of size $2 \times 26$ ($k=13$).}
\label{fig:collapse_m2NC} 
\end{figure}

Although $U^c_2$ varies with $k$, we confirm that the critical exponents 
for the N-C transition remains the same as those of the two dimensional Ising model. 
To do so, we determine the critical exponents for the system with $m=2$ and 
$k=13$ using finite size scaling. For this example, critical Binder cumulant is
$\approx 0.35$, noticeably different from that for the Ising universality
class. The data for the Binder cumulant $U_2$ for different system 
sizes intersect at $\mu_c \approx 1.00$ [see fig.~\ref{fig:collapse_m2NC} (a)]. 
We find that the data for $U_2$, $Q_2$ and $\chi_2$ for different system sizes 
collapse onto a single curve when scaled as in  Eq.~(\ref{eq:scaling}) with 
Ising exponents  
$\beta/\nu=1/8$, $\gamma/\nu=7/4$, and $\nu = 1$ 
[see Fig.~\ref{fig:collapse_m2NC} (b)--(d)]. We thus conclude that, though
the critical Binder cumulant is non-universal, the transition is in the 
Ising universality class.

\section{\label{sec:IN_MF}Estimation of the I-N phase boundary using analytical methods}

In this section we obtain the asymptotic behavior of the isotropic-nematic 
phase boundary for large $k$ using analytical methods. In the absence of
an exact solution, we present two approximate calculations: first a Bethe approximation
and second a virial expansion truncated at the second virial coefficient.

\subsection{Bethe Approximation}

The Bethe approximation becomes exact on tree like lattices. For $m=1$,
the model was solved exactly on the 4-coordinated random locally tree like 
layered lattice (RLTL) to obtain $\rho_c^{I-N} = 2/(k-1)$~\cite{dhar2011} or
$A_1=2$. The RLTL also allows an exact solution to be obtained for more
complicated systems like repulsive rods~\cite{joyjit_rltl2013}. However, a
convenient formulation of the problem of hard rectangles on the RLTL is 
lacking. Therefore, we resort to an ad-hoc Bethe approximation introduced by DiMarzio 
to estimate the entropy of hard rods on 
a cubic lattice~\cite{dimarzio1961}, and later used for
studying the statistics of hard rods on different 
lattices~\cite{fernandez2008d,fernandez2008c,linares2008}. However,
a straightforward extension of this method to a system of rectangles
suffers from the enumeration result depending on the order in which the rectangles are
placed. A scheme that overcomes this shortcoming was suggested in
Ref.~\cite{sokolova2000} and was implemented by us to study the 
N-C transition~\cite{joyjit_rectangle}. Here, we adapt the
calculations to study the I-N transition.

The I-N phase boundary can be determined if the entropy as a function of the  
densities of the horizontal and vertical rectangles is known.
We estimate the entropy by computing the number of ways of 
placing $N_x$ horizontal and $N_y$ vertical rectangles on the lattice. 

\begin{figure}
\includegraphics[width=\columnwidth]{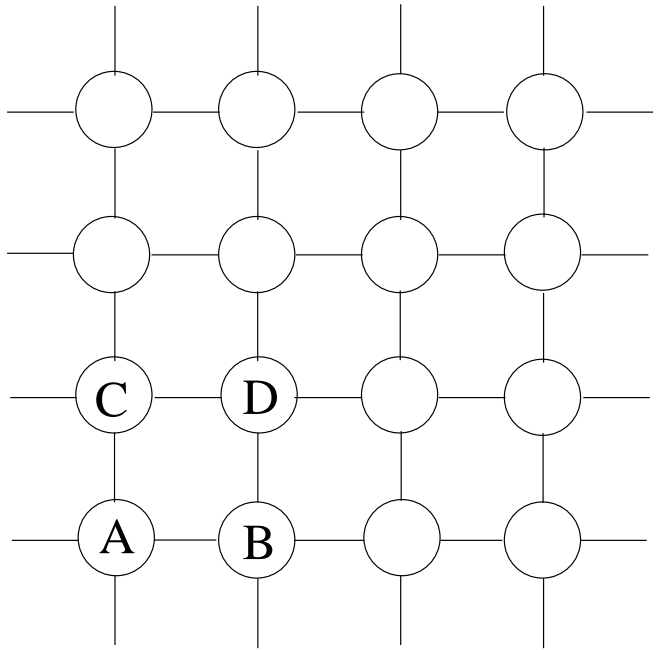}
\caption{(Color online) Schematic of a square lattice showing the position of the sites A-B to explain 
the calculation of the isotropic-nematic phase boundary.}
\label{fig:lat_in}
\end{figure}

First, we 
place the horizontal rectangles on the lattice one by one. 
Given that $j_x$ horizontal rectangles have 
been placed, the number of ways of placing the $(j_x+1)^{th}$ horizontal rectangle
may be estimated as follows.  The head of the rectangle may be placed in one of
the $(M-m^2 k j_x)$ empty sites, where $M$ is the total number of lattice sites. 
We denote this site by $A$  (see Fig.~\ref{fig:lat_in}). For this new configuration 
to be valid, all sites in the $m \times m k$ rectangle with head at $A$ should be 
empty.  Given $A$ is empty, we divide the remaining 
$(m^2 k-1)$ sites in three groups: $(mk-1)$ sites along the
line $AB$, $(m-1)$ sites along the line $AC$, and 
the remaining $(m-1) (mk-1)$ sites ($D$ is an example).  
Let $P_x(B|A)$ be the conditional probability that $B$ is empty given that $A$ is 
empty. Then the probability that $(mk-1)$ sites along the
line $AB$ are empty is $[P_x(B|A)]^{mk-1}$, where the subscript $x$ denotes
the direction $AB$. In writing this, we ignore all correlations beyond the nearest 
neighbor. Likewise, the probability that $(m-1)$ sites along the line $AC$ are empty
is given by $[P_y(C|A)]^{m-1}$, where $P_y(C|A)$ is the conditional probability that 
$C$ is empty given $A$ is empty. Let $P(D|B\cap C)$ denote the conditional 
probability that $D$ is empty given that $B$ and $C$ are both empty. Then, the
probability that the remaining $(m-1)(mk-1)$ sites are empty  may be approximated by
$[P(D|B\cap C)]^{(m-1)(mk-1)}$. Collecting these different terms together, we
obtain the number of ways to place the $(j_x+1)^{th}$ horizontal rectangle
\bea
\nu_{j_x+1} &=& (M-m^2 k j_x) \times [P_x(B|A)]^{mk-1} [P_y(C|A)]^{m-1} \nonumber \\
&\times& [P(D|B \cap C)]^{(m-1)(mk-1)}.
\label{eq:nu}
\eea

It is not possible to determine these conditional probabilities exactly. However,
they may be estimated by assuming that the rectangles are placed randomly. 
Given $A$ is empty, either $B$ might be empty or occupied by a 
horizontal rectangle (as no vertical 
rectangles have been placed yet) in $m$ ways. Thus, 
given $A$ is empty, the probability that $B$ is also empty, is 
\be
P_x(B|A)=\frac{M-m^2 k j_x}{M-m^2 k j_x+m j_x}.
\label{eq:pba}
\ee 
Similarly, if $A$ is empty, $C$ might be empty or it might be occupied by any of the $mk$ sites 
on the longer axis (passing through $C$) of a horizontal rectangle. Thus the probability that $C$ is empty, given $A$ empty is given by 
\be
P_y(C|A)=\frac{M-m^2 k j_x}{M-m^2 k j_x+m k j_x}. 
\label{eq:pca}
\ee 

Next we estimate $P(D|B\cap C)$. 
If we follow a similar approach to calculate $P(D|B \cap C)$, 
the resultant entropy becomes dependent on the
order of placement of the horizontal and vertical rectangles, and thus asymmetric 
with respect to $N_x$ and
$N_y$. To overcome this shortcoming, we follow the Bethe approximation
proposed in Ref.~\cite{sokolova2000} and assume 
\be
P(D|B \cap C) \approx \frac{P_x(C|D) P_y(B|D)}{P_{xy}(C|B)}, 
\label{eq:pdbc}
\ee
where 
\be
P_{xy}(B|C)=\frac{M-m^2 k j_x}{M-m(m-1)k j_x+(m-1) j_x},
\label{eq:pbc}
\ee 
is the 
probability that $C$ is empty given $B$ is empty. It can be easily seen that 
\begin{subequations}
\bea
P_x(C|D)=P_x(B|A), \\ 
P_y(B|D)=P_y(C|A). 
\eea
\label{eq:equality}
\end{subequations}

As all the horizontal rectangles are indistinguishable, 
the total number of ways to place $N_x$ of them is, 
\be
\Omega_x = \frac{1}{N_x!}\displaystyle\prod_{j_x=0}^{N_x-1} \nu_{j_x+1}.
\ee
Substituting Eqs.~(\ref{eq:pba})--(\ref{eq:equality}) into Eq.~(\ref{eq:nu}), we obtain 
$\nu_{j_x+1}$. $\Omega_x$ is given by 
\begin{widetext}
\bea
\Omega_x =\frac{1}{N_x!}\displaystyle\prod_{j_x=0}^{N_x-1} 
\frac{[M-m^2 k j_x]^{m^2 k}[M-(m-1)(mk-1) j_x]^{(m-1)(mk-1)}}
{[M-m (mk-1) j_x]^{m(mk-1)}[M-m k (m-1) j_x]^{m k (m-1)}}.
\label{eq:omega_x}
\eea
\end{widetext}

After placing $N_x$ horizontal rectangles we would like to determine the number of ways in which 
$N_y$ vertical rectangles may be placed on the lattice. Given $N_x$ horizontal rectangles and $j_y$ vertical rectangles 
have already been placed, we estimate $\nu_{j_y+1}$, the number of ways to place the 
$(j_y+1)^{th}$ vertical rectangle, using the same procedure as above. Now, we may choose an empty site $A$ (see Fig.~\ref{fig:lat_in}) 
randomly in $(M-m^2 k N_x-m^2 k j_y)$ ways to place the head of the 
$(j_y+1)^{th}$ vertical rectangle. As the vertical 
rectangles have their longer axis along $y$-direction, it can be easily seen that
\bea
\nu_{j_y+1} &=& (M-m^2 k N_x-m^2k j_y) [P_y(C|A)]^{mk-1} \nonumber \\
&\times&  [P_x(B|A)]^{m-1} [P(D|B \cap C)]^{(m-1)(mk-1)}.
\label{eq:nu_y}
\eea

The expressions for the conditional probabilities will now be modified due to 
the presence of both horizontal and vertical rectangles. 
If $A$ is empty, $C$ may  be empty or occupied by one of the $m k$ sites on the 
long axis (passing through $C$) of a horizontal rectangle, 
or by one of the $m$ sites on the short axis (passing through $C$) of a 
vertical rectangle. Hence, given $A$ is 
empty, the probability that $C$ is also empty is
\be
P_y(C|A)= \frac{M-m^2 k N_x-m^2 k j_y}{M-mk(m-1) N_x-m(mk-1) j_y}.
\label{eq:pca_y}
\ee

Similarly, the probability of $B$ being empty, given $A$ is empty, is
\be
P_x(B|A) = \frac{M-m^2 k N_x-m^2 k j_y}{M-m(mk-1) N_x-mk(m-1)j_y}.
\label{eq:pba_y}
\ee
Now the probability that $B$ is empty, given $C$ is empty, is 
\be
P_{xy}(B|C)=\frac{M-m^2k N_x-m^2 kj_y}{M-(mk-1)(m-1) (N_x+j_y)}.
\label{eq:pbc_y}
\ee
$P(D|B\cap C)$ is determined using Eqs.~(\ref{eq:pdbc}) and (\ref{eq:equality}). 
Substituting Eqs.~(\ref{eq:pca_y})--(\ref{eq:pbc_y}) into Eq.~(\ref{eq:nu_y}), we obtain $\nu_{j_y+1}$. 
The total number of ways to place $N_y$ vertical rectangles, given 
that $N_x$ horizontal rectangles 
have already been placed, is then
\begin{widetext}
\bea
\Omega_y = \frac{1}{N_y!}\displaystyle\prod_{j_y=0}^{N_y-1} \nu_{j_y+1}
&=&\frac{1}{N_y!}\displaystyle\prod_{j_y=0}^{N_y-1} 
\frac{[M-m^2 k (N_x+j_y)]^{m^2 k}}{[M-m(mk-1)N_x+mk(m-1)j_y]^{mk(m-1)}} 
\nonumber \\
&\times& \frac{[M-(m-1)(mk-1)N_x-(m-1)(mk-1)j_y]^{(m-1)(mk-1)}}{
[M-mk(m-1)N_x-m(mk-1)j_y]^{m(mk-1)}}.
\label{eq:omega_y}
\eea
\end{widetext}
The total number of ways to place $N_x$ horizontal and $N_y$ vertical rectangles 
on the lattice is given by
\be
\Omega=\Omega_x \Omega_y.
\ee

Let $\rho_x$ and $\rho_y$ be the fraction of the sites occupied by the horizontal 
and the vertical rectangles, given by
\be
\rho_i =  \frac{m^2 k N_i}{M},\quad i=x,y.
\ee
Using Eqs.~(\ref{eq:omega_x}) and (\ref{eq:omega_y}), the entropy of 
per site in the thermodynamic limit may be expressed in terms of $\rho_x$ and $\rho_y$ as 
\begin{widetext}
\bea
s{(\rho_x,\rho_y)} &=& \lim_{M\to\infty}\frac{1}{M} \ln 
\left(\Omega_x \Omega_y \right) \nonumber \\ 
&=&- \sum_{i=x,y}\frac{\rho_i}{m^2 k}\ln \frac{\rho_i}{m^2 k}
-\left[1-\rho\right] \ln \left[1-\rho\right] 
-\left[1-\frac{(m-1)(mk-1)}{m^2 k}\rho\right]
\ln \left[1-\frac{(m-1)(mk-1)}{m^2 k}\rho\right]
\nonumber \\ &&
+\sum_{i=x,y} \left[1-\frac{(mk-1)}{mk}\rho+\frac{(k-1)}{mk}\rho_i\right] 
\ln \left[1-\frac{(mk-1)}{mk}\rho+\frac{(k-1)}{mk}\rho_i\right],
\eea
\end{widetext}
where $\rho=\rho_x+\rho_y$ is the fraction of occupied sites.

The entropy $s(\rho_x,\rho_y)$ is not concave everywhere. The true entropy $\bar{s}(\rho_x,\rho_y)$ 
is obtained by the Maxwell construction such that
\be
\bar{s}(\rho_x,\rho_y)= \mathcal{CE} \left[ s(\rho_x,\rho_y)
\right],
\ee
where $ \mathcal{CE}$ denotes the concave envelope. 

\begin{figure}
\includegraphics[width=\columnwidth]{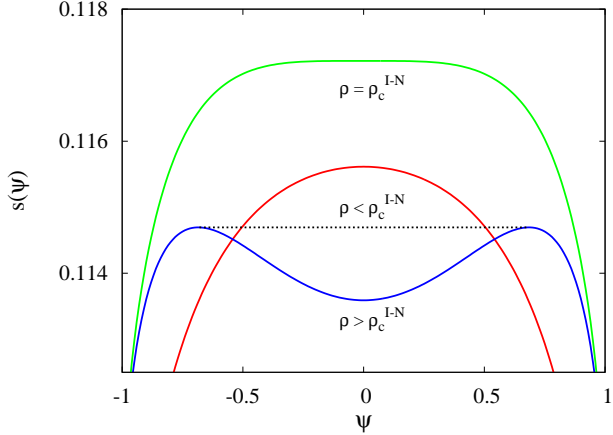}
\caption{(Color online) Entropy $s$ as a function of 
the nematic order parameter $\psi$ near the I-N transition ($\rho_c^{I-N} \approx 0.552$). 
The data are for $k=4$ and $m=2$. The dotted line denotes the concave envelope}
\label{fig:s_q_in}
\end{figure}

The entropy may also be expressed in terms of the total density $\rho=\rho_x+\rho_y$ 
and the nematic order parameter $\psi$, defined as
\be
\psi=\frac{\rho_x-\rho_y}{\rho}.
\ee
$\psi$ is zero in the isotropic phase and non-zero in the nematic phase. 
At a fixed density $\rho$, the preferred phase is 
obtained by maximizing $s(\psi)$ with respect to $\psi$. The transition density for the I-N 
transition is denoted by $\rho_c^{I-N}$.
In Fig.~\ref{fig:s_q_in} we show the plot of entropy $s(\psi)$ as a function of $\psi$, 
for three different densities near the I-N transition. For $\rho < \rho_c^{I-N}$ 
the entropy $s(\psi)$ is maximum at $\psi =0$ i.e $\rho_x=\rho_y$, corresponding 
to the isotropic phase. Beyond $\rho_c^{I-N}$ 
the entropy develops two symmetric maxima at $\psi = \pm \psi_0$, where $\psi_0=0$ at $\rho=\rho_c^{I-N}$. 
$\psi_0 \neq 0$ i.e. $\rho_x \neq \rho_y$ corresponds to the nematic phase. 
The order parameter $\psi$ grows continuously with density $\rho$.  
This is a typical signature of a continuous transition with two equivalent broken symmetry phases. 
The entropy $s(\rho,\psi)$ is invariant under the transformation 
$\psi \leftrightarrow -\psi$ and contains only even powers of $\psi$, when 
expanded about $\psi=0$. The critical density $\rho_c^{I-N}$ may be obtained by solving 
$d^2 s /d \psi^2|_{\psi=0}=0$ and is given by
\be
\rho_c^{I-N} = \frac{2 k m}{m k^2+m-k-1}.
\ee
Asymptotic behavior of $\rho_c^{I-N}$ is given by
\be
\rho_c^{I-N} = \begin{cases} \frac{2}{k}+ \frac{2}{m k^2} + O(k^{-3}), 
&k\to \infty, m \mbox{ fixed,} \\ 
\frac{2 k}{1+k^2}\! + \!\frac{2 k (1+k)}{(1+k^2)^2 m}+ O(m^{-2}),
& m \to \infty, k \mbox{ fixed}. \end{cases}
\label{eq:asym_dim}
\ee
Thus, $A_1=2$.

When $m=1$, the critical density $\rho_c^{I-N}=2/(k-1)$, which matches with the exact 
calculation of $\rho_c^{I-N}$ for the system of hard rods of length $k$ on the 
RLTL~\cite{dhar2011}. 
It reflects that the Bethe approximations becomes exact on the RLTL. 
For $m=1$, the nematic phase and hence the I-N transition exists
for $k \geq k_{min}=4$.
While for $m=2$ and $3$, 
$k_{min}=3$, for $m \geq 4$ the nematic phase exists even for $k=2$. 

\subsection{Virial Expansion}

In this subsection we determine $\rho_c^{I-N}$ using a standard virial expansion
truncated at the second virial coefficient. We closely follow the calculations
of Zwanzig for oriented hard rectangles in the continuum~\cite{zwanzig1963}.
The excess free energy of the system of hard rectangles (relative to 
the ideal gas) may be expressed in terms of the virial coefficients and the density. We truncate the 
series at the second virial coefficient and study the I-N transition in the limit $k\to \infty$. 

Consider a system of $N$ rectangles on the square lattice of volume $V$. Each rectangle may be 
oriented along two possible directions. Setting $\beta=1$, the configurational sum 
of the system is given by,
\be
Q_N=\frac{1}{N! 2^N} \sum_\textbf{u} \sum_\textbf{R} \exp(-U_N),
\label{eq:partition}
\ee
where the sum over all possible positions and directions are denoted by $\sum_\textbf{R}$ and 
$\sum_\textbf{u}$ respectively, $U_N$ is the total interaction energy of all rectangles. The 
excess free energy (relative to the ideal gas) $\phi_N$ of the system of rectangles having fixed orientations 
is defined by
\be
\exp[-\phi_N (\textbf{u})]=\frac{1}{V^N}  \sum_\textbf{R} \exp(-\beta U_N).
\label{eq:excess_free}
\ee
As the rectangles having same orientation are indistinguishable, $\phi_N$ depends only on the 
fractions of the rectangles pointing along the two possible directions. If the number of 
rectangles oriented along direction $i$ is denoted by $N_i$, we may rewrite the 
Eq.~(\ref{eq:partition}) using Eq.~(\ref{eq:excess_free}) as
\bea
Q_N &=&\frac{V^N}{N! 2^N} \sum_{N_1, N_2=0}^N \frac{N!}{N_1! N_2!} 
e^{-\phi_N (N_1,N_2)} \delta_{N_1+N_2,N}\nonumber \\
&=& \sum_{N_1=0}^N \sum_{N_2=0}^N W(N_1,N_2), 
\label{eq:final_partition}
\eea
where $\delta_{N_1+N_2,N}$ takes care of the constraint that the total number of rectangles is $N$ and $W$ is given by 
\be
W(N_1,N_2)=\frac{V^N}{2^N N_1! N_2!} \exp [-\phi_N (N_1,N_2)].
\ee
In the thermodynamic limit $N \to \infty$ and $V \to \infty$, the 
above summation may be replaced by the largest summand $W_{max}$ 
with negligible error. Thus the configurational free energy per particle is given by 
\be
f=-\lim_{N, V\to \infty} \frac{1}{N}\ln Q_N=-\lim_{N, V\to \infty} \frac{1}{N} W_{max}.
\label{eq:f_x}
\ee

The fractions of rectangles pointing in the $i$ -direction is denoted by $x_i=N_i/N$, such that $(x_1+x_2)=1$, 
and the number density of the rectangles is given by $N/V=\rho/m^2k$, 
where $\rho$ is the total fraction of occupied sites. Equation~(\ref{eq:f_x}) for the free energy may be 
expressed in terms of $x_1$ and $x_2$ as
\be
f(x_1,x_2)=-1+\ln 2+\ln \rho+\displaystyle\sum_{i=1}^2 x_i \ln x_i+\frac{1}{N}\phi_N(\rho,x_1,x_2).
\label{eq:f_rho}
\ee
The virial expansion of the excess free energy $\phi_N$, for a composition $\textbf{x}=(x_1,x_2)$ of the rectangles is given by 
\be
-\frac{1}{N} \phi_N (\rho, \textbf{x})= \displaystyle \sum_{n=2} B_n(\textbf{x}) \left (\frac{\rho}{m^2k}\right)^{n-1},
\label{eq:virial}
\ee
where 
\bea
B_n(\textbf{x})&=&\frac{1}{V n!} \int \sum \prod f \nonumber \\
&=&\frac{1}{V n!} \displaystyle\sum_{j=0}^n \binom{n}{j} x_1^{n-j} x_2^j B(n-j,j) \nonumber \\
&=& \frac{1}{V}\displaystyle\sum_{j=0}^n \frac{B(n-j,j)}{(n-j)!j!} x_1^{n-j} x_2^j,
\label{eq:def_vir}
\eea
where $\int \sum \prod f$ is the standard abbreviation for the cluster integrals over 
the irreducible graphs consist of $n$ rectangles with composition $\textbf{x}$ and $f$ denotes the Mayer functions, defined as
\be
f=\exp(-U)-1,
\ee
where $U$ is the interaction energy. Due to the hard core exclusion, we have 
$U=\infty$ for any intersection or overlap among the rectangles, otherwise $U=0$. Hence
\be
f = \begin{cases} -1, & \mbox{for any intersection} \\
  0, &\mbox{otherwise} \end{cases}
  \label{eq:mayer}
\ee
$B(n-j,j)$ denotes the sum of the irreducible $n$-particle graphs for
the composition where $(n-j)$ rectangles are oriented along
the $x$ direction and $j$ rectangles are along the $y$-direction. 

As the total fraction $x_1+x_2=1$, we set
\bea
x_1 &=& x, \nonumber \\
x_2 &=& 1-x.
\eea
We consider up to the 
second virial coefficient and truncate the expansion 
in Eq.~(\ref{eq:virial}) at first order in $\rho$. 
From the definition of the virial coefficients in Eq.~(\ref{eq:def_vir}), 
we can easily infer 
that they are symmetric in the following way:
\be
B(n_1,n_2)=B(n_2,n_1).
\label{eq:sym}
\ee
Using Eq.~(\ref{eq:def_vir}) and the above symmetry property of $B(n_1,n_2)$, 
we can rewrite Eq.~(\ref{eq:virial}) as
\bea
-\frac{1}{N} \phi_N &\approx& \frac{1}{2V} B(2,0) \left(\frac{\rho}{m^2 k}\right) (2x^2-2x+1)\nonumber \\
&+& \frac{1}{V} B(1,1) \left(\frac{\rho}{m^2k}\right) (2x-2x^2)+O(\rho^2).
\label{eq:virial02}
\eea

\begin{figure}
\includegraphics[width=\columnwidth]{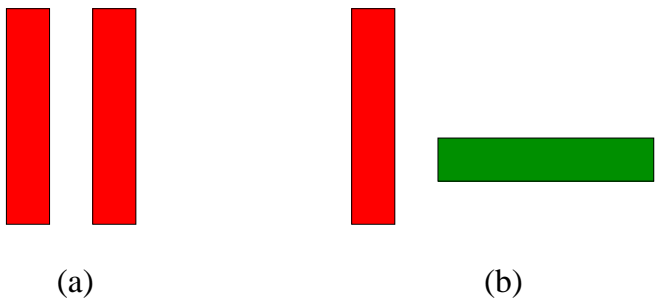}
\caption{(Color online) Schematic diagram showing the orientations of two 
rectangles in the calculation of (a) $B(2,0)$ and (b) $B(1,1)$.
}
\label{fig:zwanzig}
\end{figure}

Now we evaluate the virial coefficients. From Eq.~(\ref{eq:mayer}) we can see that 
$f$ has nonzero contributions only when the rods intersect. 
Thus the calculation of the virial coefficients on a lattice turns out as 
the problem of counting the number of disallowed configurations. 
By definition
\bea
B(2,0)&=&B(0,2)=\int d^2 R_1 \int d^2 R_2 \mbox{ } f_{12} (2,0) \nonumber \\
&=& -V\times(2mk-1)\times(2m-1),
\label{eq:b20}
\eea
where $(2mk-1)\times(2m-1)$ is the number of disallowed configurations when 
both the rectangles are oriented along the same direction 
[see Fig.~\ref{fig:zwanzig}(a)]. Similarly
\bea
B(1,1)&=&\int d^2 R_1 \int d^2 R_2 \mbox{ } f_{12} (1,1) \nonumber \\
&=&-V \times (m+mk-1)^2,
\label{eq:b11}
\eea
where $(m+mk-1)^2$ is the number of disallowed configurations when the 
two rectangles are oriented along different directions [see
Fig.~\ref{fig:zwanzig}(b)].

Substituting Eqs.~(\ref{eq:b20}) and (\ref{eq:b11}) into 
Eq.~(\ref{eq:virial02}), we find
\bea
-\frac{1}{N}\phi_N &\approx& -\frac{1}{2} (2x^2-2x+1) \rho (2m-1)(2mk-1) \nonumber \\
&-& (2x-2x^2) \rho (m+mk-1)^2+O(\rho^2).
\label{eq:phi02}
\eea
Now substituting Eq.~(\ref{eq:phi02}) in Eq.~(\ref{eq:f_rho}), the 
expression for the free energy reduces to 
\bea
f(x) &=& -1+\log 2+\log \frac{\rho}{m^2k}+x \log x+(1-x)\nonumber \\
&\times& \log (1-x) +(2x-2x^2) \rho (m+mk-1)^2 \nonumber \\
&+& (x^2-x+ \frac{1}{2}) \rho (2m-1)(2mk-1) 
+O(\rho^2).
\eea

The preferred state at any fixed density is obtained by minimizing the free energy $f(x)$ with respect to $x$. 
For $\rho<\rho_c^{I-N}$, $f(x)$ is minimized for $x=1/2$, corresponding to the isotropic phase, and 
beyond $\rho_c^{I-N}$, $f(x)$ is minimized for $x\neq 1/2$, corresponding to the nematic phase. Thus the system 
undergoes a transition from an isotropic phase to a nematic phase with increasing density.  
The I-N transition is found to be continuous with the critical density $\rho_c^{I-N}. 
$The expansion of the free energy $f(x)$ as a power series in $x$ about $x=1/2$ contains 
only even powers, and thus the critical density $\rho_c^{I-N}$ may be determined by solving
\be
\frac{d^2}{d x^2} f(x) |_{x=\frac{1}{2}}=0.
\label{eq:rhoc}
\ee
By solving Eq.~(\ref{eq:rhoc}) for $\rho$, we find 
\be
\rho_c^{I-N}=\frac{2 k m^2}{1-2(1+k)m+2(1+k^2)m^2}.
\ee

The 
asymptotic behavior of $\rho_c^{I-N}$ is given by
\be
\rho_c^{I-N} = \begin{cases} \frac{1}{k}+ \frac{1}{m k^2} + O(k^{-3}), 
&k\to \infty, m \mbox{ fixed,} \\ 
\frac{k}{1+k^2} \! + \! \frac{k (1+k)}{(1+k^2)^2 m}+ O(m^{-2}),
& m \to \infty, k \mbox{ fixed}. \end{cases}
\label{eq:asym_vir}
\ee
Comparing Eq.~(\ref{eq:asym_dim}) and Eq.~(\ref{eq:asym_vir}), we see that 
both the Bethe approximation and the virial theory predicts 
$\rho_c^{I-N}\approx A_1/k$ for $k \gg 1$. The virial calculation
gives $A_1=1$. 

\section{\label{sec:summary}Summary and Discussion}

For $k\geq 7$, the system of long, hard rectangles of size $m\times mk$ on the 
square lattice undergoes three entropy driven phase transitions with 
density: first from a low density I phase to an intermediate density N 
phase, second from the N phase to a C phase and third from the C phase 
to a high density S phase~\cite{joyjit_rectangle}. In this paper we 
study the I-N and the N-C transition when $k \gg 1$. From extensive
Monte Carlo simulations of systems with $m=1$, $2$ and $3$, we establish that 
$\rho_c^{I-N} \approx A_1/k$, for $k \gg 1$, where $A_1$ is independent 
of $m$ and is estimated to be $4.80\pm 0.05$, the numerical value
being consistent with that obtained from simulation of oriented
lines~\cite{fischer2009}.
The maximum value of $k$ studied in the paper is $60$, earlier simulations 
having  been restricted up to $m=1$ and $k=12$~\cite{fernandez2008c}. The I-N 
transition was also studied analytically using an ad-hoc Bethe 
approximation and a truncated virial expansion. Both these theories 
support the numerical result $\rho_c^{I-N} \approx A_1/k$, for $k \gg 
1$, where $A_1$ is independent of $m$. While the Bethe 
approximation gives $A_1=2$, the truncated second virial theory predicts 
$A_1=1$.

The Bethe approximation, while taking into account nearest neighbor
correlations, ignores other correlations and there appears to be 
no systematic way of improving the calculations to obtain better
estimates of $A_1$. On the other hand, the virial 
expansion truncated at the second virial coefficient is known to
become exact in  three dimensions when $k \to \infty$. 
But in two dimensions, higher 
order virial coefficients contributes significantly. To confirm this, we 
computed the higher order virial coefficients. As $B_2 \sim k^2$ [see 
Eqs.~(\ref{eq:b11})], in the limit $k\to 
\infty$, $B_2 \times \rho/k \sim O(1)$. We can rewrite 
Eq.~(\ref{eq:virial}) as
\bea
-\frac{1}{N} \phi_N (\textbf{x})&\approx& B_2(\textbf{x}) \frac{\rho}{m^2k}
+\frac{B_3(\textbf{x})}{[B_2(\textbf{x})]^2} \left[B_2(\textbf{x})\frac{\rho}{m^2 k}\right]^2\nonumber \\
&+& \frac{B_4(\textbf{x})}{[B_2(\textbf{x})]^3} 
\left[B_2(\textbf{x})\frac{\rho}{m^2 k}\right]^3 +O(\rho^4).
\eea
When $k\gg 1$, it can be verified that $B_3\sim O(k^3)$ and hence 
$B_3/[B_2]^2 \sim O(1/k)$. Quite interestingly we find $B_4 \sim 
O(k^6)$ and $B_4/[B_2]^3 \sim O(1)$. Thus $B_4$ will have non 
negligible contribution to $\rho_c^{I-N}$. 
In general $B_{2n} \sim O(k^{4n-2})$, implying all the even virial 
coefficients will have non 
negligible contributions. Usually, the number of diagrams required to
compute higher order virial coefficients increase rapidly with order.
However, here the number of diagrams are of order one. Hence, it may
be possible to determine $A_1$ exactly by taking into account
account all the even virial coefficients.

We also numerically investigated the asymptotic behavior of $\rho_c^{N-C}$ for 
$m=2$ and find $\rho_c^{N-C} \approx 0.73+0.23 k^{-1}$ when $k \gg 1$, 
which is in qualitative agreement with the prediction of the Bethe 
approximation: $\rho_c^{N-C} \approx A_2+A_3/k$, for $k \gg 1$, presented in 
Ref.~\cite{joyjit_rectangle}. For larger $m$, we expect the transition
to become first order, however the asymptotic results is likely to be
qualitatively the same. Taking the limit $k\to \infty$ keeping
$m$ fixed corresponds to a system of thin, long hard rectangles in
the continuum. Thus, we expect the N-C transition to persist in
continuum models.

Density functional theory calculations for 
a system of hard rectangles with restricted
orientation in the continuum, confined in a two dimensional square nanocavity, predicts that 
the system will exhibit nematic, smectic, columnar and solid like 
phases, where the solid-like phase has 
both orientational and complete positional order~\cite{yuri2013}. 
In contrast, we do not find any evidence of
smectic or solid-like phases when $m$ or $k$ tend to $\infty$, the
continuum limit. 
In particular, on lattices the maximal density phase of a monodispersed system does 
not have orientational order~\cite{degennesBook,ghosh2007}. 
It would therefore be important to verify the phase diagram of 
hard rectangles with restricted orientation in two dimensional
continuum through direct numerical simulations, similar to the
simulations for rectangles with continuous
orientation~\cite{bates1998,bates2000,frenkel2004,donev2006}.

We showed that the critical Binder cumulant for the N-C transition decreases 
as $k^{-1}$ with increasing the aspect ratio $k$ of the 
rectangles. The critical Binder cumulant in the Ising model on
rectangular geometry decreases as $\alpha^{-1}$, where $\alpha$ is the
aspect ratio of the lattice~\cite{blote1993}. Whether a mapping
between $k$ and $\alpha$ exists is an open question. Curiously, the
critical Binder cumulant is zero when $k \to \infty$ (or $\alpha \to
\infty$). In the Ising
model, this has been interpreted as the absence of transition on
one-dimensional geometries~\cite{blote1993}. However, the hard
rectangle system shows a transition at $k\to \infty$. It is possible
that in this limit, the fluctuations at the transition become
gaussian.

The critical density for the high-density C-S transition was argued to
be of the form $1-a/(mk^2)$ for $k \gg 1$, where $a$ is a 
constant~\cite{joyjit_rectangle}.
However, we could not numerically verify this claim
as it becomes difficult to equilibrate the system at densities close
to one due to the presence of long-lived metastable states.
Thus, the Monte Carlo algorithm 
needs further improvement. One possible direction is the modification
suggested in Ref.~\cite{kabir2014}, where fully packed 
configurations are simulated 
using transfer matrices. 

The hard rectangle model may be generalized in 
different directions.  Including attractive interaction 
results in phases with 
broken orientational and transitional
symmetry even for dimers~\cite{jesper2005,fradkin2014}. Such phases
may also be seen in mixtures of hard particles, for example dimers and
squares~\cite{kabir2014}. Another generalization is to study
polydispersed systems. In the continuum, polydispersity may result in
reentrant nematic phase or two distinct nematic
phases~\cite{sear2000,cuesta2003}. It would be interesting to see
which features persist in the lattice version of
rods~\cite{rs14} or rectangles.
These are promising areas for 
further study.

\begin{acknowledgments}
We thank W. Selke, D. Frenkel, D. Dhar and J. F. Stilck for helpful
discussions.
The simulations were carried out on the supercomputing
machine Annapurna at The Institute of Mathematical Sciences. 
\end{acknowledgments}


\begin{thebibliography}{57}%
\makeatletter
\providecommand \@ifxundefined [1]{%
 \@ifx{#1\undefined}
}%
\providecommand \@ifnum [1]{%
 \ifnum #1\expandafter \@firstoftwo
 \else \expandafter \@secondoftwo
 \fi
}%
\providecommand \@ifx [1]{%
 \ifx #1\expandafter \@firstoftwo
 \else \expandafter \@secondoftwo
 \fi
}%
\providecommand \natexlab [1]{#1}%
\providecommand \enquote  [1]{``#1''}%
\providecommand \bibnamefont  [1]{#1}%
\providecommand \bibfnamefont [1]{#1}%
\providecommand \citenamefont [1]{#1}%
\providecommand \href@noop [0]{\@secondoftwo}%
\providecommand \href [0]{\begingroup \@sanitize@url \@href}%
\providecommand \@href[1]{\@@startlink{#1}\@@href}%
\providecommand \@@href[1]{\endgroup#1\@@endlink}%
\providecommand \@sanitize@url [0]{\catcode `\\12\catcode `\$12\catcode
  `\&12\catcode `\#12\catcode `\^12\catcode `\_12\catcode `\%12\relax}%
\providecommand \@@startlink[1]{}%
\providecommand \@@endlink[0]{}%
\providecommand \url  [0]{\begingroup\@sanitize@url \@url }%
\providecommand \@url [1]{\endgroup\@href {#1}{\urlprefix }}%
\providecommand \urlprefix  [0]{URL }%
\providecommand \Eprint [0]{\href }%
\providecommand \doibase [0]{http://dx.doi.org/}%
\providecommand \selectlanguage [0]{\@gobble}%
\providecommand \bibinfo  [0]{\@secondoftwo}%
\providecommand \bibfield  [0]{\@secondoftwo}%
\providecommand \translation [1]{[#1]}%
\providecommand \BibitemOpen [0]{}%
\providecommand \bibitemStop [0]{}%
\providecommand \bibitemNoStop [0]{.\EOS\space}%
\providecommand \EOS [0]{\spacefactor3000\relax}%
\providecommand \BibitemShut  [1]{\csname bibitem#1\endcsname}%
\let\auto@bib@innerbib\@empty
\bibitem [{\citenamefont {Taylor}\ \emph {et~al.}(1985)\citenamefont {Taylor},
  \citenamefont {Williams}, \citenamefont {Park}, \citenamefont {Bartelt},\
  and\ \citenamefont {Einstein}}]{taylor1985}%
  \BibitemOpen
  \bibfield  {author} {\bibinfo {author} {\bibfnamefont {D.~E.}\ \bibnamefont
  {Taylor}}, \bibinfo {author} {\bibfnamefont {E.~D.}\ \bibnamefont
  {Williams}}, \bibinfo {author} {\bibfnamefont {R.~L.}\ \bibnamefont {Park}},
  \bibinfo {author} {\bibfnamefont {N.~C.}\ \bibnamefont {Bartelt}}, \ and\
  \bibinfo {author} {\bibfnamefont {T.~L.}\ \bibnamefont {Einstein}},\
  }\href@noop {} {\bibfield  {journal} {\bibinfo  {journal} {Phys. Rev. B}\
  }\textbf {\bibinfo {volume} {32}},\ \bibinfo {pages} {4653} (\bibinfo {year}
  {1985})}\BibitemShut {NoStop}%
\bibitem [{\citenamefont {Bak}\ \emph {et~al.}(1985)\citenamefont {Bak},
  \citenamefont {Kleban}, \citenamefont {Unertl}, \citenamefont {Ochab},
  \citenamefont {Akinci}, \citenamefont {Bartelt},\ and\ \citenamefont
  {Einstein}}]{bak1985}%
  \BibitemOpen
  \bibfield  {author} {\bibinfo {author} {\bibfnamefont {P.}~\bibnamefont
  {Bak}}, \bibinfo {author} {\bibfnamefont {P.}~\bibnamefont {Kleban}},
  \bibinfo {author} {\bibfnamefont {W.~N.}\ \bibnamefont {Unertl}}, \bibinfo
  {author} {\bibfnamefont {J.}~\bibnamefont {Ochab}}, \bibinfo {author}
  {\bibfnamefont {G.}~\bibnamefont {Akinci}}, \bibinfo {author} {\bibfnamefont
  {N.~C.}\ \bibnamefont {Bartelt}}, \ and\ \bibinfo {author} {\bibfnamefont
  {T.~L.}\ \bibnamefont {Einstein}},\ }\href@noop {} {\bibfield  {journal}
  {\bibinfo  {journal} {Phys. Rev. Lett.}\ }\textbf {\bibinfo {volume} {54}},\
  \bibinfo {pages} {1539} (\bibinfo {year} {1985})}\BibitemShut {NoStop}%
\bibitem [{\citenamefont {D\"{u}nweg}\ \emph {et~al.}(1991)\citenamefont
  {D\"{u}nweg}, \citenamefont {Milchev},\ and\ \citenamefont
  {Rikvold}}]{rikvold1991}%
  \BibitemOpen
  \bibfield  {author} {\bibinfo {author} {\bibfnamefont {B.}~\bibnamefont
  {D\"{u}nweg}}, \bibinfo {author} {\bibfnamefont {A.}~\bibnamefont {Milchev}},
  \ and\ \bibinfo {author} {\bibfnamefont {P.~A.}\ \bibnamefont {Rikvold}},\
  }\href@noop {} {\bibfield  {journal} {\bibinfo  {journal} {J. Chem. Phys.}\
  }\textbf {\bibinfo {volume} {94}},\ \bibinfo {pages} {3958} (\bibinfo {year}
  {1991})}\BibitemShut {NoStop}%
\bibitem [{\citenamefont {Koper}(1998)}]{koper1998}%
  \BibitemOpen
  \bibfield  {author} {\bibinfo {author} {\bibfnamefont {M.~T.}\ \bibnamefont
  {Koper}},\ }\href@noop {} {\bibfield  {journal} {\bibinfo  {journal} {J.
  Electroanal. Chem.}\ }\textbf {\bibinfo {volume} {450}},\ \bibinfo {pages}
  {189} (\bibinfo {year} {1998})}\BibitemShut {NoStop}%
\bibitem [{\citenamefont {Patrykiejew}\ \emph {et~al.}(2000)\citenamefont
  {Patrykiejew}, \citenamefont {Sokolowski},\ and\ \citenamefont
  {Binder}}]{binder2000}%
  \BibitemOpen
  \bibfield  {author} {\bibinfo {author} {\bibfnamefont {A.}~\bibnamefont
  {Patrykiejew}}, \bibinfo {author} {\bibfnamefont {S.}~\bibnamefont
  {Sokolowski}}, \ and\ \bibinfo {author} {\bibfnamefont {K.}~\bibnamefont
  {Binder}},\ }\href@noop {} {\bibfield  {journal} {\bibinfo  {journal} {Surf.
  Sci. Rep.}\ }\textbf {\bibinfo {volume} {37}},\ \bibinfo {pages} {207}
  (\bibinfo {year} {2000})}\BibitemShut {NoStop}%
\bibitem [{\citenamefont {Liu}\ and\ \citenamefont {Evans}(2000)}]{evans2000}%
  \BibitemOpen
  \bibfield  {author} {\bibinfo {author} {\bibfnamefont {D.-J.}\ \bibnamefont
  {Liu}}\ and\ \bibinfo {author} {\bibfnamefont {J.~W.}\ \bibnamefont
  {Evans}},\ }\href@noop {} {\bibfield  {journal} {\bibinfo  {journal} {Phys.
  Rev. B}\ }\textbf {\bibinfo {volume} {62}},\ \bibinfo {pages} {2134}
  (\bibinfo {year} {2000})}\BibitemShut {NoStop}%
\bibitem [{\citenamefont {Alder}\ and\ \citenamefont
  {Wainwright}(1957)}]{alder1957}%
  \BibitemOpen
  \bibfield  {author} {\bibinfo {author} {\bibfnamefont {B.~J.}\ \bibnamefont
  {Alder}}\ and\ \bibinfo {author} {\bibfnamefont {T.~E.}\ \bibnamefont
  {Wainwright}},\ }\href@noop {} {\bibfield  {journal} {\bibinfo  {journal} {J.
  Chem. Phys.}\ }\textbf {\bibinfo {volume} {27}},\ \bibinfo {pages} {1208}
  (\bibinfo {year} {1957})}\BibitemShut {NoStop}%
\bibitem [{\citenamefont {Alder}\ and\ \citenamefont
  {Wainwright}(1962)}]{alder1962}%
  \BibitemOpen
  \bibfield  {author} {\bibinfo {author} {\bibfnamefont {B.~J.}\ \bibnamefont
  {Alder}}\ and\ \bibinfo {author} {\bibfnamefont {T.~E.}\ \bibnamefont
  {Wainwright}},\ }\href@noop {} {\bibfield  {journal} {\bibinfo  {journal}
  {Phys. Rev.}\ }\textbf {\bibinfo {volume} {127}},\ \bibinfo {pages} {359}
  (\bibinfo {year} {1962})}\BibitemShut {NoStop}%
\bibitem [{\citenamefont {Dhar}(1982)}]{deepak2}%
  \BibitemOpen
  \bibfield  {author} {\bibinfo {author} {\bibfnamefont {D.}~\bibnamefont
  {Dhar}},\ }\href@noop {} {\bibfield  {journal} {\bibinfo  {journal} {Phys.
  Rev. Lett.}\ }\textbf {\bibinfo {volume} {49}},\ \bibinfo {pages} {959}
  (\bibinfo {year} {1982})}\BibitemShut {NoStop}%
\bibitem [{\citenamefont {Dhar}(1983)}]{deepak1}%
  \BibitemOpen
  \bibfield  {author} {\bibinfo {author} {\bibfnamefont {D.}~\bibnamefont
  {Dhar}},\ }\href {\doibase 10.1103/PhysRevLett.51.853} {\bibfield  {journal}
  {\bibinfo  {journal} {Phys. Rev. Lett.}\ }\textbf {\bibinfo {volume} {51}},\
  \bibinfo {pages} {853} (\bibinfo {year} {1983})}\BibitemShut {NoStop}%
\bibitem [{\citenamefont {Brydges}\ and\ \citenamefont
  {Imbrie}(2003)}]{imbrie}%
  \BibitemOpen
  \bibfield  {author} {\bibinfo {author} {\bibfnamefont {D.~C.}\ \bibnamefont
  {Brydges}}\ and\ \bibinfo {author} {\bibfnamefont {J.~Z.}\ \bibnamefont
  {Imbrie}},\ }\href {\doibase 10.1023/A:1022143331697} {\bibfield  {journal}
  {\bibinfo  {journal} {J. Stat. Phys.}\ }\textbf {\bibinfo {volume} {110}},\
  \bibinfo {pages} {503} (\bibinfo {year} {2003})}\BibitemShut {NoStop}%
\bibitem [{\citenamefont {Parisi}\ and\ \citenamefont
  {Sourlas}(1981)}]{parisi}%
  \BibitemOpen
  \bibfield  {author} {\bibinfo {author} {\bibfnamefont {G.}~\bibnamefont
  {Parisi}}\ and\ \bibinfo {author} {\bibfnamefont {N.}~\bibnamefont
  {Sourlas}},\ }\href {\doibase 10.1103/PhysRevLett.46.871} {\bibfield
  {journal} {\bibinfo  {journal} {Phys. Rev. Lett.}\ }\textbf {\bibinfo
  {volume} {46}},\ \bibinfo {pages} {871} (\bibinfo {year} {1981})}\BibitemShut
  {NoStop}%
\bibitem [{\citenamefont {Gaunt}\ and\ \citenamefont
  {Fisher}(1965)}]{gaunt1965}%
  \BibitemOpen
  \bibfield  {author} {\bibinfo {author} {\bibfnamefont {D.~S.}\ \bibnamefont
  {Gaunt}}\ and\ \bibinfo {author} {\bibfnamefont {M.~E.}\ \bibnamefont
  {Fisher}},\ }\href@noop {} {\bibfield  {journal} {\bibinfo  {journal} {J.
  Chem. Phys.}\ }\textbf {\bibinfo {volume} {43}},\ \bibinfo {pages} {2840}
  (\bibinfo {year} {1965})}\BibitemShut {NoStop}%
\bibitem [{\citenamefont {Bellemans}\ and\ \citenamefont
  {Nigam}(1967)}]{nigam1967}%
  \BibitemOpen
  \bibfield  {author} {\bibinfo {author} {\bibfnamefont {A.}~\bibnamefont
  {Bellemans}}\ and\ \bibinfo {author} {\bibfnamefont {R.~K.}\ \bibnamefont
  {Nigam}},\ }\href@noop {} {\bibfield  {journal} {\bibinfo  {journal} {J.
  Chem. Phys.}\ }\textbf {\bibinfo {volume} {46}},\ \bibinfo {pages} {2922}
  (\bibinfo {year} {1967})}\BibitemShut {NoStop}%
\bibitem [{\citenamefont {Pearce}\ and\ \citenamefont
  {Seaton}(1988)}]{pearce1988}%
  \BibitemOpen
  \bibfield  {author} {\bibinfo {author} {\bibfnamefont {P.~A.}\ \bibnamefont
  {Pearce}}\ and\ \bibinfo {author} {\bibfnamefont {K.~A.}\ \bibnamefont
  {Seaton}},\ }\href@noop {} {\bibfield  {journal} {\bibinfo  {journal} {J.
  Stat. Phys.}\ }\textbf {\bibinfo {volume} {53}},\ \bibinfo {pages} {1061}
  (\bibinfo {year} {1988})}\BibitemShut {NoStop}%
\bibitem [{\citenamefont {Baram}\ and\ \citenamefont
  {Fixman}(1994)}]{baram1994}%
  \BibitemOpen
  \bibfield  {author} {\bibinfo {author} {\bibfnamefont {A.}~\bibnamefont
  {Baram}}\ and\ \bibinfo {author} {\bibfnamefont {M.}~\bibnamefont {Fixman}},\
  }\href@noop {} {\bibfield  {journal} {\bibinfo  {journal} {J. Chem. Phys.}\
  }\textbf {\bibinfo {volume} {101}},\ \bibinfo {pages} {3172} (\bibinfo {year}
  {1994})}\BibitemShut {NoStop}%
\bibitem [{\citenamefont {Feng}\ \emph {et~al.}(2011)\citenamefont {Feng},
  \citenamefont {Bl\"{o}te},\ and\ \citenamefont {Nienhuis}}]{nienhuis2011}%
  \BibitemOpen
  \bibfield  {author} {\bibinfo {author} {\bibfnamefont {X.}~\bibnamefont
  {Feng}}, \bibinfo {author} {\bibfnamefont {H.~W.~J.}\ \bibnamefont
  {Bl\"{o}te}}, \ and\ \bibinfo {author} {\bibfnamefont {B.}~\bibnamefont
  {Nienhuis}},\ }\href@noop {} {\bibfield  {journal} {\bibinfo  {journal}
  {Phys. Rev. E}\ }\textbf {\bibinfo {volume} {83}},\ \bibinfo {pages} {061153}
  (\bibinfo {year} {2011})}\BibitemShut {NoStop}%
\bibitem [{\citenamefont {Ramola}\ and\ \citenamefont
  {Dhar}(2012)}]{kabir2012}%
  \BibitemOpen
  \bibfield  {author} {\bibinfo {author} {\bibfnamefont {K.}~\bibnamefont
  {Ramola}}\ and\ \bibinfo {author} {\bibfnamefont {D.}~\bibnamefont {Dhar}},\
  }\href@noop {} {\bibfield  {journal} {\bibinfo  {journal} {Phys. Rev. E}\
  }\textbf {\bibinfo {volume} {86}},\ \bibinfo {pages} {031135} (\bibinfo
  {year} {2012})}\BibitemShut {NoStop}%
\bibitem [{\citenamefont {Baxter}(1980)}]{baxter1980}%
  \BibitemOpen
  \bibfield  {author} {\bibinfo {author} {\bibfnamefont {R.~J.}\ \bibnamefont
  {Baxter}},\ }\href@noop {} {\bibfield  {journal} {\bibinfo  {journal} {J.
  Phys. A}\ }\textbf {\bibinfo {volume} {13}},\ \bibinfo {pages} {L61}
  (\bibinfo {year} {1980})}\BibitemShut {NoStop}%
\bibitem [{\citenamefont {Heilmann}\ and\ \citenamefont
  {Praestgaard}(1973)}]{heilmann1973}%
  \BibitemOpen
  \bibfield  {author} {\bibinfo {author} {\bibfnamefont {O.~J.}\ \bibnamefont
  {Heilmann}}\ and\ \bibinfo {author} {\bibfnamefont {E.}~\bibnamefont
  {Praestgaard}},\ }\href@noop {} {\bibfield  {journal} {\bibinfo  {journal}
  {J. Stat. Phys}\ }\textbf {\bibinfo {volume} {9}},\ \bibinfo {pages} {23}
  (\bibinfo {year} {1973})}\BibitemShut {NoStop}%
\bibitem [{\citenamefont {Dickman}(2012)}]{dickman2012}%
  \BibitemOpen
  \bibfield  {author} {\bibinfo {author} {\bibfnamefont {R.}~\bibnamefont
  {Dickman}},\ }\href@noop {} {\bibfield  {journal} {\bibinfo  {journal} {J.
  Chem. Phys.}\ }\textbf {\bibinfo {volume} {136}},\ \bibinfo {pages} {174105}
  (\bibinfo {year} {2012})}\BibitemShut {NoStop}%
\bibitem [{\citenamefont {Verberkmoes}\ and\ \citenamefont
  {Nienhuis}(1999)}]{nienhuis1999}%
  \BibitemOpen
  \bibfield  {author} {\bibinfo {author} {\bibfnamefont {A.}~\bibnamefont
  {Verberkmoes}}\ and\ \bibinfo {author} {\bibfnamefont {B.}~\bibnamefont
  {Nienhuis}},\ }\href@noop {} {\bibfield  {journal} {\bibinfo  {journal}
  {Phys. Rev. Lett.}\ }\textbf {\bibinfo {volume} {83}},\ \bibinfo {pages}
  {3986} (\bibinfo {year} {1999})}\BibitemShut {NoStop}%
\bibitem [{\citenamefont {Barnes}\ \emph {et~al.}(2009)\citenamefont {Barnes},
  \citenamefont {Siderius},\ and\ \citenamefont {Gelb}}]{barnes2009}%
  \BibitemOpen
  \bibfield  {author} {\bibinfo {author} {\bibfnamefont {B.~C.}\ \bibnamefont
  {Barnes}}, \bibinfo {author} {\bibfnamefont {D.~W.}\ \bibnamefont
  {Siderius}}, \ and\ \bibinfo {author} {\bibfnamefont {L.~D.}\ \bibnamefont
  {Gelb}},\ }\href@noop {} {\bibfield  {journal} {\bibinfo  {journal}
  {Langmuir}\ }\textbf {\bibinfo {volume} {25}},\ \bibinfo {pages} {6702}
  (\bibinfo {year} {2009})}\BibitemShut {NoStop}%
\bibitem [{\citenamefont {Ghosh}\ and\ \citenamefont {Dhar}(2007)}]{ghosh2007}%
  \BibitemOpen
  \bibfield  {author} {\bibinfo {author} {\bibfnamefont {A.}~\bibnamefont
  {Ghosh}}\ and\ \bibinfo {author} {\bibfnamefont {D.}~\bibnamefont {Dhar}},\
  }\href@noop {} {\bibfield  {journal} {\bibinfo  {journal} {Euro. Phys.
  Lett.}\ }\textbf {\bibinfo {volume} {78}},\ \bibinfo {pages} {20003}
  (\bibinfo {year} {2007})}\BibitemShut {NoStop}%
\bibitem [{\citenamefont {Kundu}\ \emph {et~al.}(2013)\citenamefont {Kundu},
  \citenamefont {Rajesh}, \citenamefont {Dhar},\ and\ \citenamefont
  {Stilck}}]{joyjit2013}%
  \BibitemOpen
  \bibfield  {author} {\bibinfo {author} {\bibfnamefont {J.}~\bibnamefont
  {Kundu}}, \bibinfo {author} {\bibfnamefont {R.}~\bibnamefont {Rajesh}},
  \bibinfo {author} {\bibfnamefont {D.}~\bibnamefont {Dhar}}, \ and\ \bibinfo
  {author} {\bibfnamefont {J.~F.}\ \bibnamefont {Stilck}},\ }\href@noop {}
  {\bibfield  {journal} {\bibinfo  {journal} {Phys. Rev. E}\ }\textbf {\bibinfo
  {volume} {87}},\ \bibinfo {pages} {032103} (\bibinfo {year}
  {2013})}\BibitemShut {NoStop}%
\bibitem [{\citenamefont {Kundu}\ and\ \citenamefont
  {Rajesh}(2014)}]{joyjit_rectangle}%
  \BibitemOpen
  \bibfield  {author} {\bibinfo {author} {\bibfnamefont {J.}~\bibnamefont
  {Kundu}}\ and\ \bibinfo {author} {\bibfnamefont {R.}~\bibnamefont {Rajesh}},\
  }\href@noop {} {\bibfield  {journal} {\bibinfo  {journal} {Phys. Rev. E}\
  }\textbf {\bibinfo {volume} {89}},\ \bibinfo {pages} {052124} (\bibinfo
  {year} {2014})}\BibitemShut {NoStop}%
\bibitem [{\citenamefont {Fernandes}\ \emph {et~al.}(2007)\citenamefont
  {Fernandes}, \citenamefont {Arenzon},\ and\ \citenamefont
  {Levin}}]{fernandes2007}%
  \BibitemOpen
  \bibfield  {author} {\bibinfo {author} {\bibfnamefont {H.~C.~M.}\
  \bibnamefont {Fernandes}}, \bibinfo {author} {\bibfnamefont {J.~J.}\
  \bibnamefont {Arenzon}}, \ and\ \bibinfo {author} {\bibfnamefont
  {Y.}~\bibnamefont {Levin}},\ }\href@noop {} {\bibfield  {journal} {\bibinfo
  {journal} {J. Chem. Phys.}\ }\textbf {\bibinfo {volume} {126}},\ \bibinfo
  {pages} {114508} (\bibinfo {year} {2007})}\BibitemShut {NoStop}%
\bibitem [{\citenamefont {Nath}\ and\ \citenamefont
  {Rajesh}(2014)}]{trisha_knn}%
  \BibitemOpen
  \bibfield  {author} {\bibinfo {author} {\bibfnamefont {T.}~\bibnamefont
  {Nath}}\ and\ \bibinfo {author} {\bibfnamefont {R.}~\bibnamefont {Rajesh}},\
  }\href@noop {} {\bibfield  {journal} {\bibinfo  {journal} {Phys. Rev. E}\
  }\textbf {\bibinfo {volume} {90}},\ \bibinfo {pages} {012120} (\bibinfo
  {year} {2014})}\BibitemShut {NoStop}%
\bibitem [{\citenamefont {Ramola}\ \emph {et~al.}(2014)\citenamefont {Ramola},
  \citenamefont {Damle},\ and\ \citenamefont {Dhar}}]{kabir2014}%
  \BibitemOpen
  \bibfield  {author} {\bibinfo {author} {\bibfnamefont {K.}~\bibnamefont
  {Ramola}}, \bibinfo {author} {\bibfnamefont {K.}~\bibnamefont {Damle}}, \
  and\ \bibinfo {author} {\bibfnamefont {D.}~\bibnamefont {Dhar}},\ }\href@noop
  {} {\bibfield  {journal} {\bibinfo  {journal} {arXiv preprint
  arXiv:1408.4943}\ } (\bibinfo {year} {2014})}\BibitemShut {NoStop}%
\bibitem [{\citenamefont {de~Gennes}\ and\ \citenamefont
  {Prost}(1995)}]{degennesBook}%
  \BibitemOpen
  \bibfield  {author} {\bibinfo {author} {\bibfnamefont {P.~G.}\ \bibnamefont
  {de~Gennes}}\ and\ \bibinfo {author} {\bibfnamefont {J.}~\bibnamefont
  {Prost}},\ }\href@noop {} {\emph {\bibinfo {title} {The Physics of Liquid
  Crystals}}}\ (\bibinfo  {publisher} {Oxford University Press},\ \bibinfo
  {address} {Oxford},\ \bibinfo {year} {1995})\ pp.\ \bibinfo {pages}
  {59--66}\BibitemShut {NoStop}%
\bibitem [{\citenamefont {Disertori}\ and\ \citenamefont
  {Giuliani}(2013)}]{giuliani2013}%
  \BibitemOpen
  \bibfield  {author} {\bibinfo {author} {\bibfnamefont {M.}~\bibnamefont
  {Disertori}}\ and\ \bibinfo {author} {\bibfnamefont {A.}~\bibnamefont
  {Giuliani}},\ }\href@noop {} {\bibfield  {journal} {\bibinfo  {journal}
  {Commun. Math. Phys.}\ }\textbf {\bibinfo {volume} {323}},\ \bibinfo {pages}
  {143} (\bibinfo {year} {2013})}\BibitemShut {NoStop}%
\bibitem [{\citenamefont {Matoz-Fernandez}\ \emph
  {et~al.}(2008{\natexlab{a}})\citenamefont {Matoz-Fernandez}, \citenamefont
  {Linares},\ and\ \citenamefont {Ramirez-Pastor}}]{fernandez2008a}%
  \BibitemOpen
  \bibfield  {author} {\bibinfo {author} {\bibfnamefont {D.~A.}\ \bibnamefont
  {Matoz-Fernandez}}, \bibinfo {author} {\bibfnamefont {D.~H.}\ \bibnamefont
  {Linares}}, \ and\ \bibinfo {author} {\bibfnamefont {A.~J.}\ \bibnamefont
  {Ramirez-Pastor}},\ }\href@noop {} {\bibfield  {journal} {\bibinfo  {journal}
  {Euro. Phys. Lett}\ }\textbf {\bibinfo {volume} {82}},\ \bibinfo {pages}
  {50007} (\bibinfo {year} {2008}{\natexlab{a}})}\BibitemShut {NoStop}%
\bibitem [{\citenamefont {Kundu}\ and\ \citenamefont
  {Rajesh}(2013)}]{joyjit_rltl2013}%
  \BibitemOpen
  \bibfield  {author} {\bibinfo {author} {\bibfnamefont {J.}~\bibnamefont
  {Kundu}}\ and\ \bibinfo {author} {\bibfnamefont {R.}~\bibnamefont {Rajesh}},\
  }\href@noop {} {\bibfield  {journal} {\bibinfo  {journal} {Phys. Rev. E}\
  }\textbf {\bibinfo {volume} {88}},\ \bibinfo {pages} {012134} (\bibinfo
  {year} {2013})}\BibitemShut {NoStop}%
\bibitem [{\citenamefont {Onsager}(1949)}]{onsager1949}%
  \BibitemOpen
  \bibfield  {author} {\bibinfo {author} {\bibfnamefont {L.}~\bibnamefont
  {Onsager}},\ }\href@noop {} {\bibfield  {journal} {\bibinfo  {journal} {Ann.
  N.Y. Acad. Sci.}\ }\textbf {\bibinfo {volume} {51}},\ \bibinfo {pages} {627}
  (\bibinfo {year} {1949})}\BibitemShut {NoStop}%
\bibitem [{\citenamefont {Zwanzig}(1963)}]{zwanzig1963}%
  \BibitemOpen
  \bibfield  {author} {\bibinfo {author} {\bibfnamefont {R.}~\bibnamefont
  {Zwanzig}},\ }\href@noop {} {\bibfield  {journal} {\bibinfo  {journal} {J.
  Chem. Phys.}\ }\textbf {\bibinfo {volume} {39}},\ \bibinfo {pages} {1714}
  (\bibinfo {year} {1963})}\BibitemShut {NoStop}%
\bibitem [{\citenamefont {Vroege}\ and\ \citenamefont
  {Lekkerkerker}(1992)}]{vroege1992}%
  \BibitemOpen
  \bibfield  {author} {\bibinfo {author} {\bibfnamefont {G.~J.}\ \bibnamefont
  {Vroege}}\ and\ \bibinfo {author} {\bibfnamefont {H.~N.~W.}\ \bibnamefont
  {Lekkerkerker}},\ }\href@noop {} {\bibfield  {journal} {\bibinfo  {journal}
  {Rep. Prog. Phys.}\ }\textbf {\bibinfo {volume} {55}},\ \bibinfo {pages}
  {1241} (\bibinfo {year} {1992})}\BibitemShut {NoStop}%
\bibitem [{\citenamefont {Fischer}\ and\ \citenamefont
  {Vink}(2009)}]{fischer2009}%
  \BibitemOpen
  \bibfield  {author} {\bibinfo {author} {\bibfnamefont {T.}~\bibnamefont
  {Fischer}}\ and\ \bibinfo {author} {\bibfnamefont {R.~L.~C.}\ \bibnamefont
  {Vink}},\ }\href@noop {} {\bibfield  {journal} {\bibinfo  {journal} {Euro.
  Phys. Lett.}\ }\textbf {\bibinfo {volume} {85}},\ \bibinfo {pages} {56003}
  (\bibinfo {year} {2009})}\BibitemShut {NoStop}%
\bibitem [{\citenamefont {Matoz-Fernandez}\ \emph
  {et~al.}(2008{\natexlab{b}})\citenamefont {Matoz-Fernandez}, \citenamefont
  {Linares},\ and\ \citenamefont {Ramirez-Pastor}}]{fernandez2008c}%
  \BibitemOpen
  \bibfield  {author} {\bibinfo {author} {\bibfnamefont {D.~A.}\ \bibnamefont
  {Matoz-Fernandez}}, \bibinfo {author} {\bibfnamefont {D.~H.}\ \bibnamefont
  {Linares}}, \ and\ \bibinfo {author} {\bibfnamefont {A.~J.}\ \bibnamefont
  {Ramirez-Pastor}},\ }\href@noop {} {\bibfield  {journal} {\bibinfo  {journal}
  {J. Chem. Phys.}\ }\textbf {\bibinfo {volume} {128}},\ \bibinfo {pages}
  {214902} (\bibinfo {year} {2008}{\natexlab{b}})}\BibitemShut {NoStop}%
\bibitem [{\citenamefont {Kundu}\ \emph {et~al.}(2012)\citenamefont {Kundu},
  \citenamefont {Rajesh}, \citenamefont {Dhar},\ and\ \citenamefont
  {Stilck}}]{joyjit_dae}%
  \BibitemOpen
  \bibfield  {author} {\bibinfo {author} {\bibfnamefont {J.}~\bibnamefont
  {Kundu}}, \bibinfo {author} {\bibfnamefont {R.}~\bibnamefont {Rajesh}},
  \bibinfo {author} {\bibfnamefont {D.}~\bibnamefont {Dhar}}, \ and\ \bibinfo
  {author} {\bibfnamefont {J.~F.}\ \bibnamefont {Stilck}},\ }\href@noop {}
  {\bibfield  {journal} {\bibinfo  {journal} {AIP Conf. Proc.}\ }\textbf
  {\bibinfo {volume} {1447}},\ \bibinfo {pages} {113} (\bibinfo {year}
  {2012})}\BibitemShut {NoStop}%
\bibitem [{\citenamefont {Kamieniarz}\ and\ \citenamefont
  {Bl\"{o}te}(1993)}]{blote1993}%
  \BibitemOpen
  \bibfield  {author} {\bibinfo {author} {\bibfnamefont {G.}~\bibnamefont
  {Kamieniarz}}\ and\ \bibinfo {author} {\bibfnamefont {H.~W.~J.}\ \bibnamefont
  {Bl\"{o}te}},\ }\href@noop {} {\bibfield  {journal} {\bibinfo  {journal} {J.
  Phys. A}\ }\textbf {\bibinfo {volume} {26}},\ \bibinfo {pages} {201}
  (\bibinfo {year} {1993})}\BibitemShut {NoStop}%
\bibitem [{\citenamefont {Selke}\ and\ \citenamefont
  {Shchur}(2005)}]{selke2005}%
  \BibitemOpen
  \bibfield  {author} {\bibinfo {author} {\bibfnamefont {W.}~\bibnamefont
  {Selke}}\ and\ \bibinfo {author} {\bibfnamefont {L.~N.}\ \bibnamefont
  {Shchur}},\ }\href@noop {} {\bibfield  {journal} {\bibinfo  {journal} {J.
  Phys. A}\ }\textbf {\bibinfo {volume} {38}},\ \bibinfo {pages} {L239–L744}
  (\bibinfo {year} {2005})}\BibitemShut {NoStop}%
\bibitem [{\citenamefont {Selke}(2007)}]{selke2007}%
  \BibitemOpen
  \bibfield  {author} {\bibinfo {author} {\bibfnamefont {W.}~\bibnamefont
  {Selke}},\ }\href@noop {} {\bibfield  {journal} {\bibinfo  {journal} {J.
  Stat. Mech.}\ ,\ \bibinfo {pages} {P04008}} (\bibinfo {year}
  {2007})}\BibitemShut {NoStop}%
\bibitem [{\citenamefont {Dhar}\ \emph {et~al.}(2011)\citenamefont {Dhar},
  \citenamefont {Rajesh},\ and\ \citenamefont {Stilck}}]{dhar2011}%
  \BibitemOpen
  \bibfield  {author} {\bibinfo {author} {\bibfnamefont {D.}~\bibnamefont
  {Dhar}}, \bibinfo {author} {\bibfnamefont {R.}~\bibnamefont {Rajesh}}, \ and\
  \bibinfo {author} {\bibfnamefont {J.~F.}\ \bibnamefont {Stilck}},\
  }\href@noop {} {\bibfield  {journal} {\bibinfo  {journal} {Phys. Rev. E}\
  }\textbf {\bibinfo {volume} {84}},\ \bibinfo {pages} {011140} (\bibinfo
  {year} {2011})}\BibitemShut {NoStop}%
\bibitem [{\citenamefont {DiMarzio}(1961)}]{dimarzio1961}%
  \BibitemOpen
  \bibfield  {author} {\bibinfo {author} {\bibfnamefont {E.}~\bibnamefont
  {DiMarzio}},\ }\href@noop {} {\bibfield  {journal} {\bibinfo  {journal} {J.
  Chem. Phys.}\ }\textbf {\bibinfo {volume} {35}},\ \bibinfo {pages} {658}
  (\bibinfo {year} {1961})}\BibitemShut {NoStop}%
\bibitem [{\citenamefont {Centres}\ and\ \citenamefont
  {Ramirez-Pastor}(2009)}]{fernandez2008d}%
  \BibitemOpen
  \bibfield  {author} {\bibinfo {author} {\bibfnamefont {P.}~\bibnamefont
  {Centres}}\ and\ \bibinfo {author} {\bibfnamefont {A.}~\bibnamefont
  {Ramirez-Pastor}},\ }\href@noop {} {\bibfield  {journal} {\bibinfo  {journal}
  {Physica A}\ }\textbf {\bibinfo {volume} {388}},\ \bibinfo {pages}
  {2001–2019} (\bibinfo {year} {2009})}\BibitemShut {NoStop}%
\bibitem [{\citenamefont {Linares}\ \emph {et~al.}(2008)\citenamefont
  {Linares}, \citenamefont {Rom\'{a}},\ and\ \citenamefont
  {Ramirez-Pastor}}]{linares2008}%
  \BibitemOpen
  \bibfield  {author} {\bibinfo {author} {\bibfnamefont {D.~H.}\ \bibnamefont
  {Linares}}, \bibinfo {author} {\bibfnamefont {F.}~\bibnamefont {Rom\'{a}}}, \
  and\ \bibinfo {author} {\bibfnamefont {A.~J.}\ \bibnamefont
  {Ramirez-Pastor}},\ }\href@noop {} {\bibfield  {journal} {\bibinfo  {journal}
  {J. Stat. Mech.}\ ,\ \bibinfo {pages} {P03013}} (\bibinfo {year}
  {2008})}\BibitemShut {NoStop}%
\bibitem [{\citenamefont {Sokolova}\ and\ \citenamefont
  {Tumanyan}(2000)}]{sokolova2000}%
  \BibitemOpen
  \bibfield  {author} {\bibinfo {author} {\bibfnamefont {E.~P.}\ \bibnamefont
  {Sokolova}}\ and\ \bibinfo {author} {\bibfnamefont {N.~P.}\ \bibnamefont
  {Tumanyan}},\ }\href@noop {} {\bibfield  {journal} {\bibinfo  {journal} {Liq.
  Crys.}\ }\textbf {\bibinfo {volume} {27}},\ \bibinfo {pages} {813} (\bibinfo
  {year} {2000})}\BibitemShut {NoStop}%
\bibitem [{\citenamefont {Gonz\'{a}lez-Pinto}\ \emph
  {et~al.}(2013)\citenamefont {Gonz\'{a}lez-Pinto}, \citenamefont
  {Mart\'{i}nez-Rat\'{o}n},\ and\ \citenamefont {Velasco}}]{yuri2013}%
  \BibitemOpen
  \bibfield  {author} {\bibinfo {author} {\bibfnamefont {M.}~\bibnamefont
  {Gonz\'{a}lez-Pinto}}, \bibinfo {author} {\bibfnamefont {Y.}~\bibnamefont
  {Mart\'{i}nez-Rat\'{o}n}}, \ and\ \bibinfo {author} {\bibfnamefont
  {E.}~\bibnamefont {Velasco}},\ }\href@noop {} {\bibfield  {journal} {\bibinfo
   {journal} {Phys. Rev. E}\ }\textbf {\bibinfo {volume} {88}},\ \bibinfo
  {pages} {032506} (\bibinfo {year} {2013})}\BibitemShut {NoStop}%
\bibitem [{\citenamefont {Bates}\ and\ \citenamefont
  {Frenkel}(1998)}]{bates1998}%
  \BibitemOpen
  \bibfield  {author} {\bibinfo {author} {\bibfnamefont {M.~A.}\ \bibnamefont
  {Bates}}\ and\ \bibinfo {author} {\bibfnamefont {D.}~\bibnamefont
  {Frenkel}},\ }\href@noop {} {\bibfield  {journal} {\bibinfo  {journal} {J.
  Chem. Phys}\ }\textbf {\bibinfo {volume} {109}},\ \bibinfo {pages} {6193}
  (\bibinfo {year} {1998})}\BibitemShut {NoStop}%
\bibitem [{\citenamefont {Bates}\ and\ \citenamefont
  {Frenkel}(2000)}]{bates2000}%
  \BibitemOpen
  \bibfield  {author} {\bibinfo {author} {\bibfnamefont {M.~A.}\ \bibnamefont
  {Bates}}\ and\ \bibinfo {author} {\bibfnamefont {D.}~\bibnamefont
  {Frenkel}},\ }\href@noop {} {\bibfield  {journal} {\bibinfo  {journal} {J.
  Chem. Phys}\ }\textbf {\bibinfo {volume} {112}},\ \bibinfo {pages} {10034}
  (\bibinfo {year} {2000})}\BibitemShut {NoStop}%
\bibitem [{\citenamefont {Wojciechowski}\ and\ \citenamefont
  {Frenkel}(2004)}]{frenkel2004}%
  \BibitemOpen
  \bibfield  {author} {\bibinfo {author} {\bibfnamefont {K.~W.}\ \bibnamefont
  {Wojciechowski}}\ and\ \bibinfo {author} {\bibfnamefont {D.}~\bibnamefont
  {Frenkel}},\ }\href@noop {} {\bibfield  {journal} {\bibinfo  {journal}
  {Comput. Methods Sci. Tech.}\ }\textbf {\bibinfo {volume} {10}},\ \bibinfo
  {pages} {235} (\bibinfo {year} {2004})}\BibitemShut {NoStop}%
\bibitem [{\citenamefont {Donev}\ \emph {et~al.}(2006)\citenamefont {Donev},
  \citenamefont {Burton}, \citenamefont {Stillinger},\ and\ \citenamefont
  {Torquato}}]{donev2006}%
  \BibitemOpen
  \bibfield  {author} {\bibinfo {author} {\bibfnamefont {A.}~\bibnamefont
  {Donev}}, \bibinfo {author} {\bibfnamefont {J.}~\bibnamefont {Burton}},
  \bibinfo {author} {\bibfnamefont {F.~H.}\ \bibnamefont {Stillinger}}, \ and\
  \bibinfo {author} {\bibfnamefont {S.}~\bibnamefont {Torquato}},\ }\href@noop
  {} {\bibfield  {journal} {\bibinfo  {journal} {Phys. Rev. B}\ }\textbf
  {\bibinfo {volume} {73}},\ \bibinfo {pages} {054109} (\bibinfo {year}
  {2006})}\BibitemShut {NoStop}%
\bibitem [{\citenamefont {Alet}\ \emph {et~al.}(2005)\citenamefont {Alet},
  \citenamefont {Jacobsen}, \citenamefont {Misguich}, \citenamefont {Pasquier},
  \citenamefont {Mila},\ and\ \citenamefont {Troyer}}]{jesper2005}%
  \BibitemOpen
  \bibfield  {author} {\bibinfo {author} {\bibfnamefont {F.}~\bibnamefont
  {Alet}}, \bibinfo {author} {\bibfnamefont {J.~L.}\ \bibnamefont {Jacobsen}},
  \bibinfo {author} {\bibfnamefont {G.}~\bibnamefont {Misguich}}, \bibinfo
  {author} {\bibfnamefont {V.}~\bibnamefont {Pasquier}}, \bibinfo {author}
  {\bibfnamefont {F.}~\bibnamefont {Mila}}, \ and\ \bibinfo {author}
  {\bibfnamefont {M.}~\bibnamefont {Troyer}},\ }\href@noop {} {\bibfield
  {journal} {\bibinfo  {journal} {Phys. Rev. Lett}\ }\textbf {\bibinfo {volume}
  {94}},\ \bibinfo {pages} {235702} (\bibinfo {year} {2005})}\BibitemShut
  {NoStop}%
\bibitem [{\citenamefont {Papanikolaou}\ \emph {et~al.}(2014)\citenamefont
  {Papanikolaou}, \citenamefont {Charrier},\ and\ \citenamefont
  {Fradkin}}]{fradkin2014}%
  \BibitemOpen
  \bibfield  {author} {\bibinfo {author} {\bibfnamefont {S.}~\bibnamefont
  {Papanikolaou}}, \bibinfo {author} {\bibfnamefont {D.}~\bibnamefont
  {Charrier}}, \ and\ \bibinfo {author} {\bibfnamefont {E.}~\bibnamefont
  {Fradkin}},\ }\href@noop {} {\bibfield  {journal} {\bibinfo  {journal} {Phys.
  Rev. B}\ }\textbf {\bibinfo {volume} {89}},\ \bibinfo {pages} {035128}
  (\bibinfo {year} {2014})}\BibitemShut {NoStop}%
\bibitem [{\citenamefont {Clarke}\ \emph {et~al.}(2000)\citenamefont {Clarke},
  \citenamefont {Cuesta}, \citenamefont {Sear}, \citenamefont {Sollich},\ and\
  \citenamefont {Speranza}}]{sear2000}%
  \BibitemOpen
  \bibfield  {author} {\bibinfo {author} {\bibfnamefont {N.}~\bibnamefont
  {Clarke}}, \bibinfo {author} {\bibfnamefont {J.~A.}\ \bibnamefont {Cuesta}},
  \bibinfo {author} {\bibfnamefont {R.}~\bibnamefont {Sear}}, \bibinfo {author}
  {\bibfnamefont {P.}~\bibnamefont {Sollich}}, \ and\ \bibinfo {author}
  {\bibfnamefont {A.}~\bibnamefont {Speranza}},\ }\href@noop {} {\bibfield
  {journal} {\bibinfo  {journal} {J. Chem. Phys}\ }\textbf {\bibinfo {volume}
  {113}},\ \bibinfo {pages} {5817} (\bibinfo {year} {2000})}\BibitemShut
  {NoStop}%
\bibitem [{\citenamefont {Mart\'{i}nez-Rat\'{o}n}\ and\ \citenamefont
  {Cuesta}(2003)}]{cuesta2003}%
  \BibitemOpen
  \bibfield  {author} {\bibinfo {author} {\bibfnamefont {Y.}~\bibnamefont
  {Mart\'{i}nez-Rat\'{o}n}}\ and\ \bibinfo {author} {\bibfnamefont {J.~A.}\
  \bibnamefont {Cuesta}},\ }\href@noop {} {\bibfield  {journal} {\bibinfo
  {journal} {J. Chem. Phys}\ }\textbf {\bibinfo {volume} {118}},\ \bibinfo
  {pages} {10164} (\bibinfo {year} {2003})}\BibitemShut {NoStop}%
\bibitem [{\citenamefont {Rajesh}\ and\ \citenamefont {Stilck}(2014)}]{rs14}%
  \BibitemOpen
  \bibfield  {author} {\bibinfo {author} {\bibfnamefont {R.}~\bibnamefont
  {Rajesh}}\ and\ \bibinfo {author} {\bibfnamefont {J.~F.}\ \bibnamefont
  {Stilck}},\ }\href@noop {} {\enquote {\bibinfo {title} {Polydispersed rods on
  the square lattice},}\ } (\bibinfo {year} {2014}),\ \bibinfo {note} {in
  preparation}\BibitemShut {NoStop}%
\end{thebibliography}

%

\end{document}